\documentclass[preprint,aps]{revtex4}
\usepackage{hyperref}
\usepackage{graphicx}
\usepackage{dcolumn}
\usepackage{bm}
\usepackage{multirow}
\usepackage{amssymb}
\usepackage{amsmath}
\usepackage{longtable}
\usepackage[utf8]{inputenc}

\begin{document}


\title{Alter-magnetic properties in the  perovskite compounds}

%
%
%
\author{
  Sining Zhang, Zhengxuan Wang, Minping Zhang, \\
  Xilin Zhang\footnotemark[1] \ and Guangtao Wang\footnotemark[2]
}
\date{}
\renewcommand{\thefootnote}{\fnsymbol{footnote}} 

\affiliation{%
  College of Physics, Henan Normal University, \\
  Xinxiang, Henan 453007, People's Republic of China%
}

\date{\today}
             
\begin{abstract}

Our high-throughput computational screening of the Materials Project database has identified 140 candidate materials exhibiting antiferromagnetic behavior in the $Pnma$ space group. Through systematic density functional theory calculations comparing various magnetic configurations, we demonstrate that 91 of these compounds preferentially stabilize in altermagnetic ground states, with 20 adopting the perovskite structure. Using NaCoF$_3$ as a prototypical example, we perform a comprehensive investigation of its electronic structure, magnetic ordering, and orbital configurations. Our analysis reveals that the magnetic Co$^{2+}$ ions occupy the 4c Wyckoff position and support three distinct antiferromagnetic (AFM) phases, among which the $G$-AFM configuration emerges as the ground state. Detailed symmetry analysis uncovers that the inter-sublattice coupling in the $G$-AFM phase is mediated by the combined time-reversal and rotation operation $\hat{T}\{R_{2x}|\frac{1}{2}\frac{1}{2}0\}$. This unique symmetry protection gives rise to several remarkable physical phenomena: (i) anisotropic optical conductivity, (ii) prominent anomalous transport effects including the anomalous Hall effect (AHE), Nernst effect (ANE), and thermal Hall effect (ATHE), and (iii) strong magneto-optical responses manifested through both Kerr and Faraday effects.

\end{abstract}

\maketitle

\footnotetext[1]{\href{zhangxilin@htu.edu.cn}{zhangxilin@htu.edu.cn}}
\footnotetext[2]{\href{wangtao@htu.cn}{wangtao@htu.cn}}


\section{Introduction}
Altermagnetism, a recently identified class of collinear magnetism, possessing both key features of antiferromagnets and ferromagnets, caused wildly interesting in the field of the condenser maters~\cite{Altermagnetic1,Altermagnetic2,yao3}. Despite the  vanishing net magnetization in altermagnetism,  they host pronounced anomalous transport phenomena~\cite{ruo2ahc1,ruo2ahc2,ANE,ATHE,AHE1,AHE2,AHE3,AHE4,AHE5,AHE6,AHE7,AHE8,yaoAHC,yaoMOKE}, such as the anomalous Hall conductivity~\cite{ANE,ATHE,AHE1,AHE2,AHE3,AHE4,AHE5,AHE6,AHE7,AHE8,yaoAHC,yaoMOKE}, Nernst effect~\cite{ANE,yaoAHC}, and thermal Hall effects~\cite{ATHE,yaoAHC}, along with magneto-optical responses including the Kerr and Faraday effects~\cite{yaoMOKE}.Traditionally, these effects were thought to require ferromagnetism, incompatible with antiferromagnets’ zero net magnetization. However, altermagnetic materials possess a unique symmetry where the connecting operation between the two magnetic sublattices is not the conventional space-inversion ($\hat{P}$) combined with time-reversal ($\hat{T}$), but rather a rotational or mirror symmetry operation coupled with time-reversal (e.g., $C_4\hat{T}$ or $M\hat{T}$)~\cite{Altermagnetic1,Altermagnetic2,yao3,TSC1}. This special symmetry results in  anisotropic charge distributions in  the real-space and  band splitting in the momentum-space, distinguishing altermagnets from conventional antiferromagnets~\cite{Weyl,Topandoptical,MnTe-ARPES,shc1}.

In altermagnets, magnetic ions with opposite spin orientations  exhibit distinct ligand-field environments from surrounding anions, which are symmetry-related through rotational or mirror operations~\cite{Altermagnetic1,Altermagnetic2}. For transition-metal oxides/fluorides with partially filled $d$-electron states, this configuration naturally combines Jahn-Teller (JT) distortions and orbital ordering~\cite{YVO,YVO-OPT,CSRO,CSRO-PRB}. When such JT-active ions with antiparallel spin alignment form exchange-coupled pairs, they create the fundamental prerequisites for altermagnetism: spin-dependent band splitting with $k$-dependent spin polarization~\cite{alter-orbital-order,alt-layer,hjzhang,shc1,giant-splitting}. So, it is proposed that many transition metal compounds would exhibit altermagnetic properties~\cite{Krempasky2024-Nature,Mazin-2023}.

The most definitive approach to verify altermagnetism consists of two essential experimental demonstrations: (i) neutron diffraction confirming the antiferromagnetic order, and (ii) spin-resolved angle-resolved photoemission spectroscopy (SR-ARPES) revealing the $k$-dependent spin splitting in the  band structure~\cite{MnTe-ARPES,RuO2-nosplitt}. However, these measurements face significant experimental challenges, particularly the demanding requirements for both high energy and spatial resolution. These challenges may lead to divergent or even conflicting conclusions among different research groups studying the same material. For instance, some studies~\cite{yaoMOKE,yaoAHC,spin-charge} have reported alter-magnetic spin moments and spin-splitting band structure in RuO$_2$, whereas others~\cite{nomagnetic,absenceAFM} have observed neither signatures of long-range magnetic order nor  spin-splitting band.

Consequently, researchers often employ indirect probes of altermagnetic properties, including: Anomalous Hall effect measurements~\cite{AHE2,AHE3,AHE4,AHE5,AHE6,AHE7,AHE8,yaoAHC},  Magneto-optical Kerr effect (MOKE)~\cite{yaoMOKE}, Faraday rotation~\cite{yaoMOKE,Chen_2024-Mn-rev} measurements. All these properties require broken P$\hat{T}$-symmetry, as they would otherwise vanish identically~\cite{zhoutong,ssg,Chen_2024-Mn-rev,mode-top}. The anomalous Hall effect exhibits strong sensitivity to crystalline symmetry operations, which are intrinsically related to magnetic configurations. This connection enables the indirect probing of the magnetic structure through such measurement. For instance, in RuO\(_2\) under \(M_{001}\hat{T} \) symmetry, the AHC component $\sigma_{xy}$ is constrained to be zero~\cite{ruo2ahc1,ruo2ahc2,RuO2-nosplitt}.

Many perovskite compounds exhibiting Jahn-Teller (JT) distortions develop concomitant magnetic ordering through JT-induced orbital ordering. A prototypical example occurs in materials with space group  \textit{Pnma},   where magnetic ions occupy 4b or 4c Wyckoff positions with alternating spin alignment (two $\uparrow$ and two $\downarrow$ sites) . Such configuration  necessarily produces altermagnetism~\cite{yao3,Krempasky2024-Nature,orthorhombic,alter-orbital-order}. Taking NaCoF$_3$ as an example, we  systematically investigate its orbital ordering, band spin-splitting, AHC, ANC, TAHC and Magneto-optical Kerr and Faraday effects~\cite{eqat1,eqat2,eqat3}.

\section{method and details}
The perovskite-structured compound NaCoF$_3$ crystallizes in the orthorhombic space group \textit{Pnma} (No.62)~\cite{orthorhombic}, with lattice parameters a=5.478\AA, b=5.665\AA, c=7.877\AA. The atomic Wyckoff positions are: Na at 4c (0.5148, 0.5545, 0.25), Co at 4b (0.5, 0.0, 0.0), F$_1$ at 8d (0.1957, 0.1996, 0.5555), and F$_2$ at 4c (0.1056, 0.4608, 0.25), as shown in the Fig.1. Each Co$^{2+}$  sites in the center of CoF$_6$ octahedra, which has different tilt angle $\varphi$ from $c$-axis and rotation angle $\theta$ around $c$-axis, maned as Jahn-Teller distortion. Such distortion may induce orbital  and magnetic orders: ferromagnetic and three anti-ferromagnetic states schematically shown in Fig.1a. In order to determine the true ground state, we have calculated the total energy of all the four different cases. The calculations were done with the BSTATE~\cite{BSTATE} code, in the  ultra-soft pseudopotential plane wave method~\cite{vasp1,vasp2}. All the lattice constants and the atomic positions  adopted in our calculations came from experimental~\cite{NaCoF3,NaCoF3-mag}. After carefully checking the convergence of calculated results with respect to the
cutoff energy and the number of $k$-points, we adopted a cutoff energy
of 30 Ry  and Monkhorst-Pack $k$-points generated
with  $17\times17\times13$. Due to the  localized nature of Co-3d electronic states in the material, we employed the GGA+U method~\cite{ggapbe,HU,ldau} with parameter values U=3.0 eV and J=1.0 eV. These parameters are consistent with those commonly used in previous calculations for Co-F compounds ~\cite{CoF1,CoF2}.
The direct observations of anisotropic Fermi surface and spin-splitting band structure are difficult. However the measurements of anisotropic optical conductivity (AOC), anomalous Hall, Nernst, and thermal Hall effects, along with magneto-optical responses including the Kerr and Faraday effects have been proved as a useful way to investigate the characters of alter-magnetism  indirectly~\cite{Chen_2024-Mn-rev,layer-nernst}. These properties   are calculated from the converged Kohn-Sham wave functions $|\psi_{n\bf k}\rangle$ and eigenvalues
$E_n({\bf k})$ by using the following Kubo formula~\cite{opteq1,opteq2,KCrF}:

\begin{eqnarray}
\sigma_{\alpha\beta}(\omega)
&=&-\frac{16}{V}\sum_{\bf k\it n}if_{n\bf k}\sum_{m}
\frac{1}{\omega_{mn}^2-(\omega+i\delta)^2} \nonumber\\
&&\left[\frac{\omega+i\delta}{\omega_{mn}}
Re(\pi_{nm}^\alpha\pi_{mn}^\beta)+iIm(\pi_{nm}^\alpha\pi_{mn}^\beta)
\right]
\end{eqnarray}
where $\alpha$ and $\beta$ (=$x,y,z$) are indices for directions,
$\omega$ is the excitation energy, $V$ is the volume of the unit cell,
$n$ and $m$ are band indices, $f_{n\bf k}$ is the Fermi distribution
function, $\omega_{mn}=E_{m}({\rm k})-E_{n}({\rm k})$ and $\delta$ is
the lifetime broadening ($\delta$=0.01Ry in this work),
$\pi_{nm}^\alpha=\langle \psi_{n\bf k}|(-i\nabla_\alpha)|\psi_{m\rm
k}\rangle$ are the matrix elements of the momentum operator.

From the above calculated $\pi_{nm}^\alpha$ and $\pi_{nm}^\beta$,  we get Berry curvature $\Omega_n^\gamma(\mathbf{k})$,  AHC, ANC and ATHC by the formulas \cite{eqat1,eqat2,eqat3} as:
\begin{eqnarray}
\Omega_n^\gamma(\mathbf{k}) &=& -2 \, \text{Im} \sum_{m \neq n} \frac{\pi_{nm}^\alpha \pi_{mn}^\beta}{(\epsilon_m - \epsilon_n)^2}   \label{Berry}\\
\sigma_{\alpha\beta}^{\mathrm{AHC}} &=& - \frac{e^2}{\hbar}\int \frac{d^3\bm{k}}{(2\pi)^3}  \Omega^{\gamma}(\bm{k}) \label{AHC} \\
\alpha_{\alpha\beta}^{\mathrm{ANC}} &=& \int \frac{\varepsilon - \mu}{eT} \left( -\frac{\partial f}{\partial \varepsilon} \right) \sigma_{\alpha\beta}^{\mathrm{AHC}} d\varepsilon \label{ANC} \\
\kappa_{\alpha\beta}^{\mathrm{ATHC}} &=& \int \frac{(\varepsilon - \mu)^2}{e^2T} \left( -\frac{\partial f}{\partial \varepsilon} \right) \sigma_{\alpha\beta}^{\mathrm{AHC}} d\varepsilon \label{ATHC}
\end{eqnarray}

where \( \Omega^{\gamma}(\bm{k}) \) is the momentum-resolved Berry curvature, \( \mu \) is the chemical potential, and \( f = 1/[\exp \left( (\varepsilon - \mu) / k_B T \right) + 1 ]\) is the Fermi-Dirac distribution function.

For the complex Kerr and Faraday angle, we adopt a simplified expression under the assumption of a small rotation angle~\cite{mokeeq1,mokeeq2,mokeeq3,mokeeq4,mokeeq5,mokeeq6,mokeeq7,mokeeq8,mokeeq9}: 
\begin{eqnarray}
    \phi^{z}_{K} = \vartheta^{z}_{K} + i\varepsilon^{z}_{K} \approx \frac{-\nu_{xyz} \sigma_{xy}}{\sigma_{0} \sqrt{1 + i(4\pi/\omega)\sigma_{0}}} ,  \\
    \phi^{z}_{F} = \vartheta^{z}_{F} + i\varepsilon^{z}_{K} \approx \frac{-\nu_{xyz} \sigma_{xy}}{ \sqrt{1 + i(4\pi/\omega)\sigma_{0}}} \frac{2\pi}{c} , 
\end{eqnarray}
where $\sigma_0 = (\sigma_{xx} +\sigma_{yy})/2$.

\begin{figure}[htbp]
\includegraphics[clip,scale=0.45]{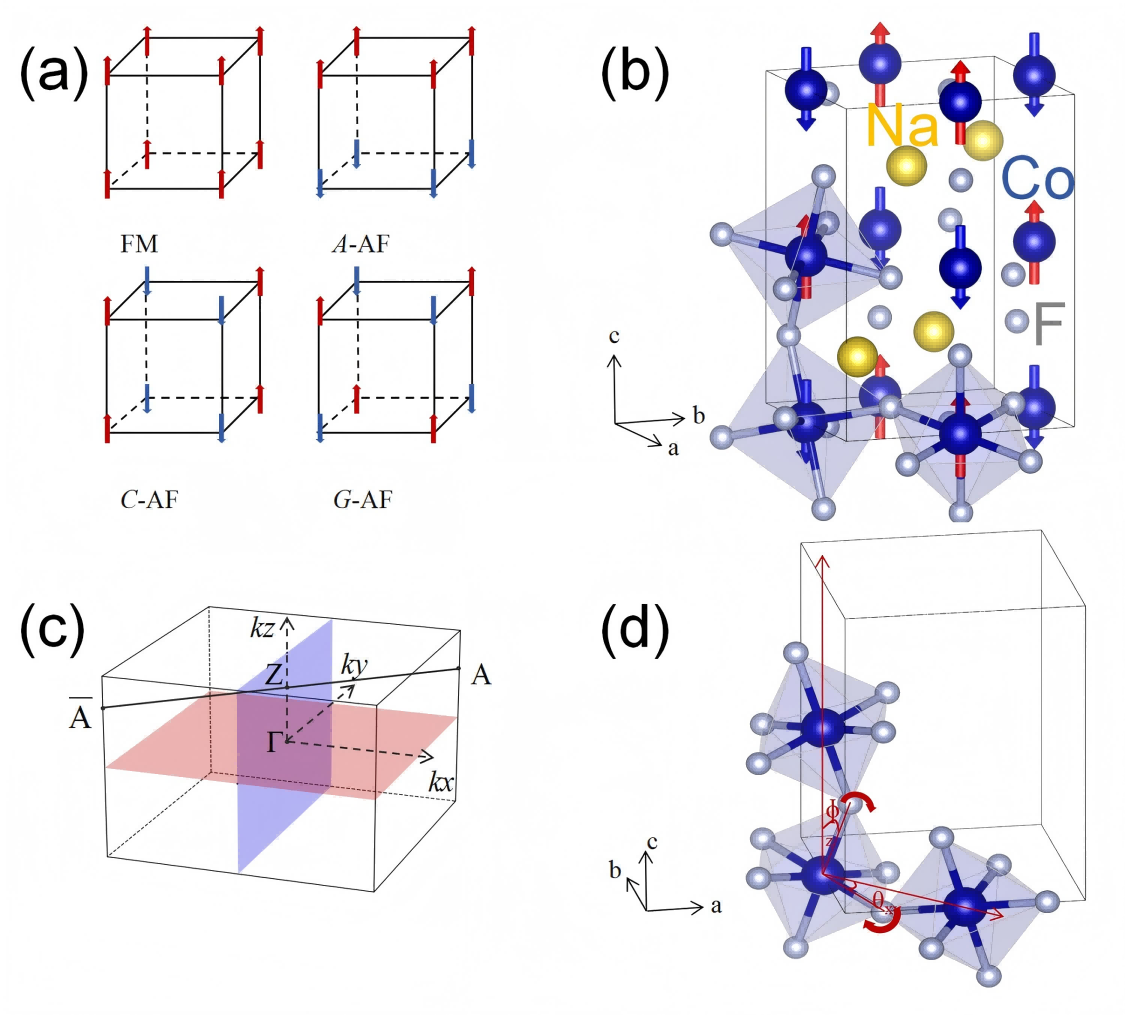}
\caption{The magnetic configurations and crystal structure of NaCoF$_3$. (a) presents the ferromagnetic and three antiferromagnetic states. (b) and (d) show the CoF$_6$  octahedra and their rotation angle $\theta$ and tilt angle $\varphi$.  (c) presents the Brillouin zone, high symmetry points and lines for the band structure calculations. }
\end{figure}

\section{Results and Discussion}
Table I presents the calculated properties of A-, C-, and G-type antiferromagnetic (AFM) states, including their total energies relative to the ferromagnetic (FM) state, in-plane (J${ab}$) and out-of-plane (J${c}$) exchange constants, Co$^{2+}$ magnetic moments, and band gaps for the G-AFM state as functions of the Hubbard parameter U${eff}$. Our calculations reveal that all three AFM configurations are more stable than the FM state, with the G-AFM configuration being the most energetically favorable. The positive values obtained for both J${ab}$ and J${c}$ are consistent with the G-AFM ground state. Notably, increasing U${eff}$ leads to: (1) a reduction in exchange coupling constants, (2) an enhancement of the magnetic moment, and (3) an increase in the band gap - in agreement with our previous findings~\cite{KCrF} on KCrF$_3$. Having established the G-AFM state as the ground state through total energy calculations, we focus our main text discussion on this magnetic configuration, while presenting results for other magnetic structures in the Supplementary Material.

\begin{table}[t]
\caption{The total energy of  $A$,  $C$, $G$-type
antiferromagnetism (with respect to FM), the exchange constant in $a-b$ plane (J$_{ab}$) and along $c$-axis J$_{c}$,
the Moment of Co ion and the band gap of $G$-AFM state, varying with  the correlation (U$_{eff}$).  $J_c=[E(F)-E(G)-E(A)+E(C)]/(4S^2)$ and $J_{ab}=[E(F)-E(G)+E(A)-E(C)]/(8S^2)$, where E($F$), E($A$), E($C$), and E($G$) were the total energy of FM, $A$-type, $C$-type and $G$-type anti-ferromagnetism.
 }
\begin{tabular}{| c| c | c | c | c | c | c |c |}
\cline{1-8}\hline\hline
       U$_{eff}$         &$A$(meV)          &$C$(meV)     &$G$(meV)      &J$_{ab}$(meV)  &J$_{c}$(meV) &Moment($\mu_{B}$)  &gap( eV)     \\
       \hline
      0 &-216 &-381  &-434  & 9.55 &8.6 &2.61 &0.45 \\
      1 &-106  &-239  &-286  & 6.68 &4.9 &2.67 &1.27 \\
      2 & -52  &-145  &-191  & 4.52 &3.1 &2.72 &2.15 \\
      3 & -41  &-113  &-150  & 3.53 &2.5 &2.76 &2.97 \\
      4 & -32  & -87  & -98  & 2.43 &1.4 &2.79 &3.79 \\
      5 & -24  & -66  & -76  & 1.88 &1.1 &2.82 &4.52 \\
    \hline
\hline
\end{tabular}
\end{table}

\begin{table}[h]
\caption{
The compound crystallizes in the orthorhombic \textit{Pnma} space group (No.~62), whose crystallographic symmetry includes eight symmetry operations as enumerated in the first row ($G$). For the three distinct antiferromagnetic (AFM) configurations, these symmetry operations can be systematically categorized into two classes: (1) The first part of four symmetry operations ($G-A$) preserve the spin channel, acting solely within a single spin subsystem.  (2) The remaining four operations ($\hat{T}A$) describe spin-flip processes mediated by the time-reversal operator $\hat{T}$, which transforms up-spin states into their down-spin counterparts and vice versa.
 }
\centering
\begin{tabular}{|l|}
\hline
\begin{tabular}{@{}l@{}}
 \quad Group $G$:  $E$, $I$, $\{R_{2x}|\frac{1}{2}\frac{1}{2}0\}$, $\{R_{2y}|\frac{1}{2}\frac{1}{2}\frac{1}{2}\}$,$\{R_{2z}|00\frac{1}{2}\}$,  $\{M_{x}|\frac{1}{2}\frac{1}{2}0\}$, $\{M_{y}|\frac{1}{2}\frac{1}{2}\frac{1}{2}\}$, $\{M_{z}|00\frac{1}{2}\}$ \\
\end{tabular} \\ \hline
\begin{tabular}{|c|c|c}
AFM   & $G-A$             & $\hat{T}A$ \\ \hline
 $A$-AFM     & $E$, $I$, $\{R_{2x}|\frac{1}{2}\frac{1}{2}0\}$, $\{M_{x}|\frac{1}{2}\frac{1}{2}0\}$,
 & $\hat{T}\{R_{2y}|\frac{1}{2}\frac{1}{2}\frac{1}{2}\}$, $\hat{T}\{M_{y}|\frac{1}{2}\frac{1}{2}\frac{1}{2}\}$, $\hat{T}\{R_{2z}|00\frac{1}{2}\}$ ,$\hat{T}\{M_{z}|00\frac{1}{2}\}$ \\ \hline
 $C$-AFM     & $E$, $I$, $\{R_{2z}|00\frac{1}{2}\}$ ,$\{M_{z}|00\frac{1}{2}\}$
& $\hat{T}\{R_{2x}|\frac{1}{2}\frac{1}{2}0\}$, $\hat{T}\{R_{2y}|\frac{1}{2}\frac{1}{2}\frac{1}{2}\}$,$\hat{T}\{M_{x}|\frac{1}{2}\frac{1}{2}0\}$, $\hat{T}\{M_{y}|\frac{1}{2}\frac{1}{2}\frac{1}{2}\}$\\\hline
$G$-AFM     &$E$, $I$, $\{R_{2y}|\frac{1}{2}\frac{1}{2}\frac{1}{2}\}$, $\{M_{y}|\frac{1}{2}\frac{1}{2}\frac{1}{2}\}$
&$\hat{T}\{R_{2x}|\frac{1}{2}\frac{1}{2}0\}$, $\hat{T}\{M_{x}|\frac{1}{2}\frac{1}{2}0\}$, $\hat{T}\{R_{2z}|00\frac{1}{2}\}$,  $\hat{T}\{M_{z}|00\frac{1}{2}\}$      \\ \hline
\end{tabular} \\
\hline
\end{tabular}
\end{table}

In the Fig.2, the spin-up and spin-down bands are exact degenerated because of    the  $\hat{T}\{R_{2z}|00\frac{1}{2}\}$ symmetry.  Since $\hat{T}\{R_{2z}|00\frac{1}{2}\} \psi (s,k_x,k_y,k_z)= \psi (-s,  k_x, k_y, -k_z)$, and  $k_z$=$-k_z$ in the the  $k_z$=0($\frac{\pi}{2}$)  plane. So the bands spin-splitting on the $k_z$=0($\frac{\pi}{2}$) plane is absent. For the similar reason, the bands on the $k_x$=0($\frac{\pi}{2}$) plane are degenerated because of the symmetry $\hat{T}\{R_{2x}|\frac{1}{2}\frac{1}{2}0\}$. In the range -4.0 to -2.0 eV, the bands are mainly derived from F-2p states, while the Co-t$_{2g}$ orbitals form the bands around the Fermi level.  Around 0.35 eV, there is a flat-band derived from one of  the Co-t$_{2g}$ orbitals, which induced orbital ordering ( as shown in Fig.5). The Co-e$_{g}$ orbitals form the bands 1.0 eV  above the Fermi level. 


The magnetic moment of the system originates predominantly from Co$^{2+}$ ions, which induces significant spin-splitting in the Co-3d-derived electronic bands. Remarkably, this spin-splitting is completely suppressed in the $k_z=0$ ($\pi/2$) plane due to symmetry protection by the combined $\hat{T}{R_{2z}|00\frac{1}{2}}$ operation. This protection mechanism is lifted along the AZ$\bar{A}$ high-symmetry line (Fig. 1c), where characteristic band splitting of approximately 20 meV emerges in the antiferromagnetic state, as clearly demonstrated in Figs. 3a-c.

\begin{figure}[htbp]
\begin{center} 
\includegraphics[clip,scale=0.4]{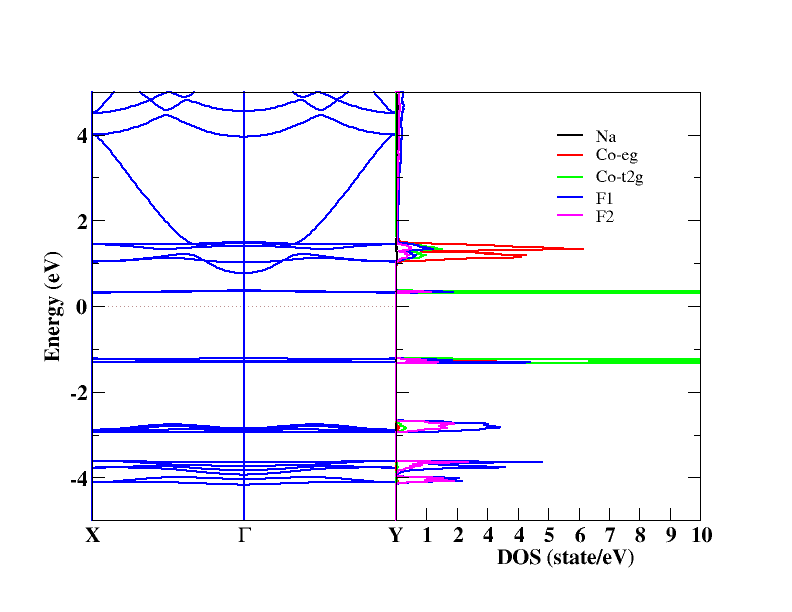}
\caption{Along the high-symmetry lines in the $k_z = 0$ plane, the calculated band structure of the $G$-AFM is presented in (a). The exact degeneracy between spin-up and spin-down channels is protected by the combined  $\hat{T}\{R_{2z}|00\frac{1}{2}\}$ symmetry. The corresponding projected density of states (PDOS) in (b) demonstrates that the electronic states near the Fermi level are predominantly contributed by the Co-$t_{2g}$ orbitals.}
\end{center}
\end{figure}

%
%

Due to spin-momentum locking in altermagnets, Fermi surface topologies can be classified as $d$-wave (e.g., RuO$_2$)~\cite{shc1,shc2}, $g$-wave (e.g., CrSb, MnTe)~\cite{giant-splitting,Chen_2024-Mn-rev,Krempasky2024-Nature}, or $i$-wave (e.g., 2H-FBr$_3$)~\cite{yao3}. Although pristine NaCoF$_3$ is semiconducting with no Fermi surface, hole doping can lower the Fermi level to reveal measurable Fermi surfaces~\cite{FS}. As shown in Fig.~4, we calculate the Fermi surface topology at $E_F = -1.25$ eV, demonstrating that $G$-type antiferromagnetic NaCoF$_3$ exhibits a $d$-wave topology similar to RuO$2$~\cite{shc1,shc2}, which arises from the nonsymmorphic symmetry operation $\hat{T}\{R_{2z}|00\frac{1}{2}\}$ coupling the two magnetic sublattices.

The Fermi surface manifests two distinctive features: (i) a pronounced $d$-wave topology~\cite{Altermagnetic1,Altermagnetic2} (Figs.4c-d), and (ii) anisotropic band splitting absent in both $k_z$=0 (Fig.4a) and $k_x$=0 (Fig.4b) planes, but clearly present in the $k_y$=0 plane (Fig.4c). The band degeneracy in the $k_z$ and $k_x$=0 planes is protected by $\hat{T}\{R_{2z}|00\frac{1}{2}\}$  and  $\hat{T}\{R_{2x}|\frac{1}{2}\frac{1}{2}0\}$,  respectively. Band splitting emerges when deviating from the $k_z$=0 plane, breaking the $\hat{T}\{R_{2z}|00\frac{1}{2}\}$ symmetry and lifting spin degeneracy, as evident from the spin-splitting FS shown in Fig.4d.

\begin{figure}[htbp]
	\begin{center}
		\includegraphics[clip,scale=0.4]{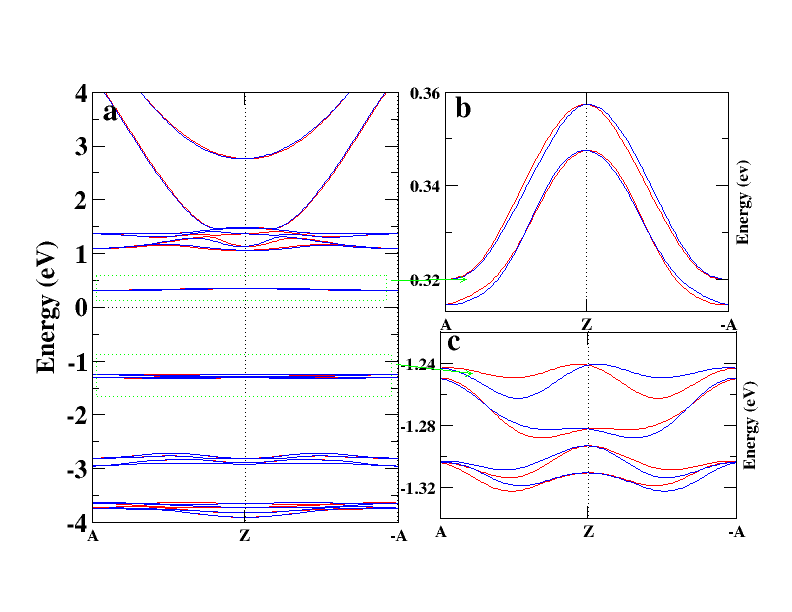}
		\caption{ Figure (a) displays the band structure calculated along the general $k$-path in the  $k_z$=0.3 plane as shown  in the Brillouin zone Fig.1(c), where the spin-up and spin-down bands are not degenerated. The splitting of the band around 0.35 eV and -1.28 eV are presented in figure (b) and (c).
		}
	\end{center}
\end{figure}

\begin{figure}[htbp]
	\begin{center} 
		\includegraphics[clip,scale=0.4]{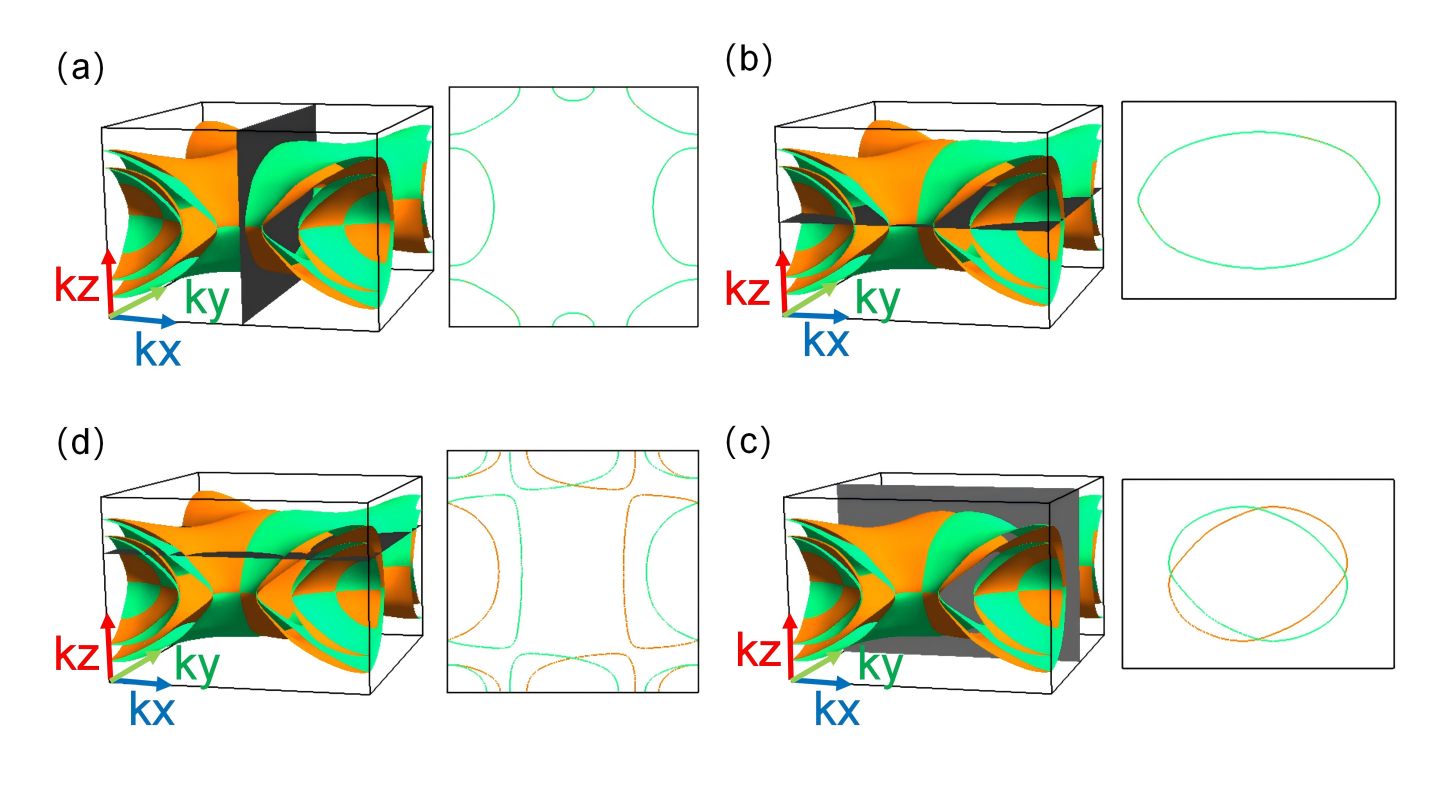}
		\caption{(a), (b)  and (c) are the Fermi surfaces (set Fermi level to -1.25 eV) and their cut-sections at  the $k_z$=0, $k_x$=0 and $k_y$=0 plane.  (d)  is the surface and its cut-section at the $k_z$=0.2. 
		}
	 \end{center}
\end{figure}

As previously discussed, the crystal field in CoF$_6$ octahedra splits the  Co-$3d$ orbitals into  higher-energy $e_g$  and lower-energy $t_{2g}$ states. Most remarkably, four $t_{2g}$ states  form four flat bands at 0.35 eV above Fermi level (Fig.2 and Fig.3), indicating very localized charge distribution. In order to study  the  charge distribution of the mentioned flat bands, we present the spin-resolved charge density  in Fig.5.  In  $G$-AFM NaCoF$_3$, the Co$^{2+}$ ions occupy the 4b Wyckoff position, yielding four magnetically inequivalent cobalt sites. Crucially, nearest-neighbor Co$^{2+}$ ions exhibit not only antiparallel spin alignment but also opposing orbital ordering patterns Fig.5a. Their magnetic and orbital patterns are closely related each other by spin space group~\cite{KCrF}. For example, Co$_1$ and Co$_3$ can be mapped to each other by the symmetry $\hat{T}\{R_{2z}|00\frac{1}{2}\}$  as shown in Fig.5a and Fig.5b.  Similarly, the symmetry $\hat{T}\{R_{2x}|\frac{1}{2}\frac{1}{2}0\}$ combines Co$_1$ with Co$_2$, as shown in Fig.5a and Fig.5c.

\begin{figure}[htbp]
	\begin{center} 
		\includegraphics[clip,scale=0.4]{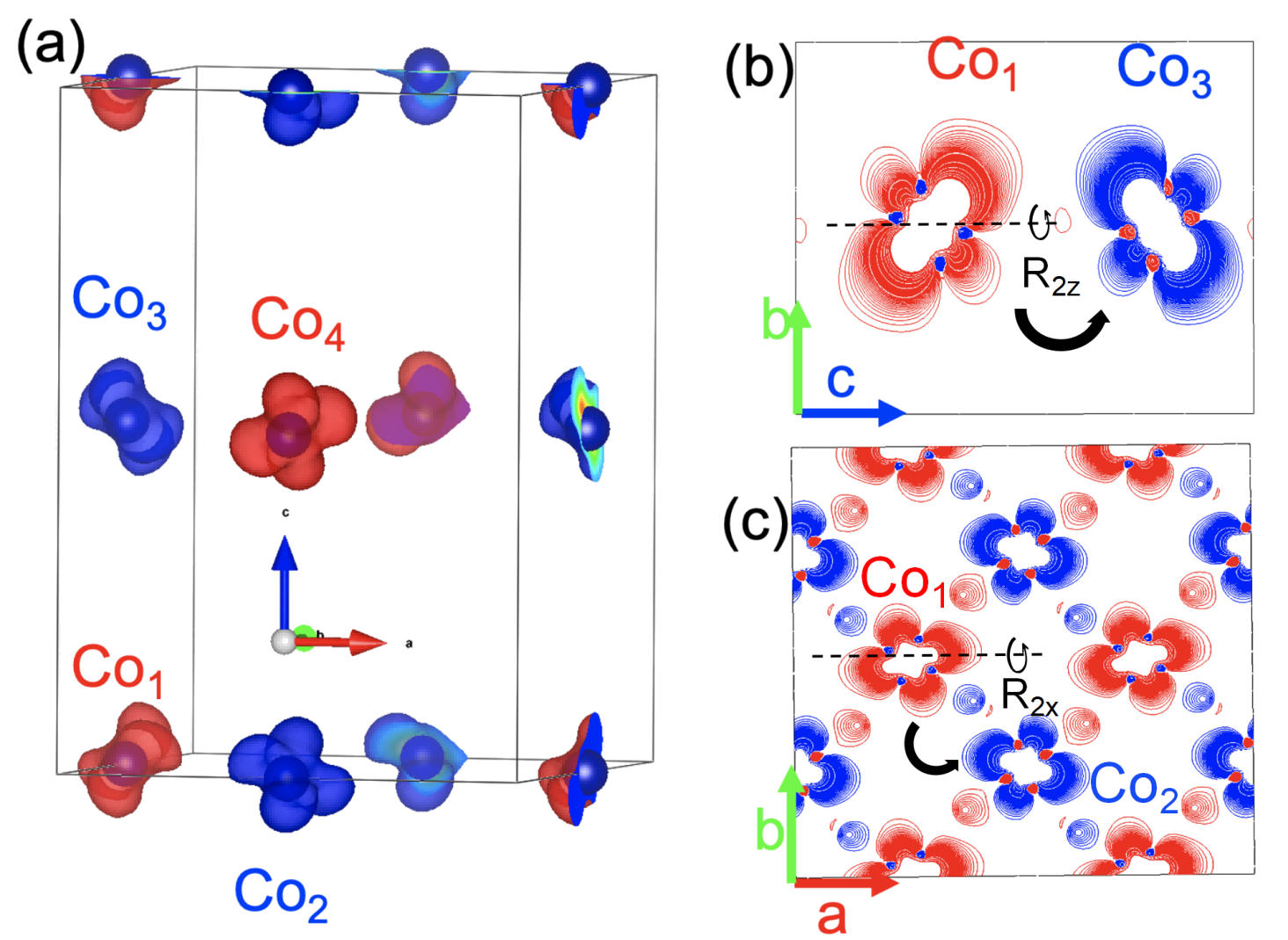}
		\caption{ The charge density of $G$-AF state, with their energy around 0.35 eV (the flat band just above the Fermi energy). (a) is the 3D charge density, while (b) and (c) are 2D charge density on the 100-plane and 001-plane, respectively.  Red  for up-spin, blue for down-spin.
		 }	
		\end{center}
\end{figure}

As we mentioned before, the direct observation of anisotropic Fermi surface and spin-splitting band structure is difficult. However such anisotropic properties can be indirectly investigated by the projected density of states of three different Co$^{2+}$ ions and optical conductivity as shown in Fig.6~\cite{KCrF,YVO-OPT,Topandoptical,Magneto}. In the a-b plane, the optical conductivity $\sigma_{\parallel}$ has two peaks, with  $\alpha_{1}$ at  3.1 eV  and $\alpha_{2}$ at about 4.3 eV. The peak $\alpha_{1}$ derived from electrons
hopping between occupied e$^{\uparrow}_{g}$ of  Co$_{1}$ and unoccupied d$^{\uparrow}_{zx}$ of Co$_{2}$.  Along c-axis,  the peak $\beta_{1}$ derived from electrons 
hopping between occupied e$^{\uparrow}_{g}$ of  Co$_{1}$ and unoccupied d$^{\uparrow}_{yz}$ of Co$_{3}$. 
The peak $\alpha_{2}$ ($\beta_{2}$ ) comes from the electrons hopping between  e$^{\uparrow}_{g}$ of Co$_{1}$ and unoccupied e$^{\uparrow}_{g}$ of Co$_{2}$ ( Co$_{3}$).  Such electronic hopping is visualized with black arrows (in the $ab$-plane) and red arrows (along the $c$-axis) in Fig.6. 

By correlating the charge density distribution presented in Fig.5 with the optical conductivity-PDOS relationship shown in Fig.6, we find that the magnetic ordering is accompanied by orbital ordering as our previous work~\cite{KCrF}.  Importantly, this established correlation provides a viable approach for detecting altermagnetic anisotropy through precise measurements of optical conductivity peak responses~\cite{KCrF,YVO-OPT,Topandoptical,Moptical}.

\begin{figure}[htbp]
	\begin{center} 
		\includegraphics[clip,scale=0.4]{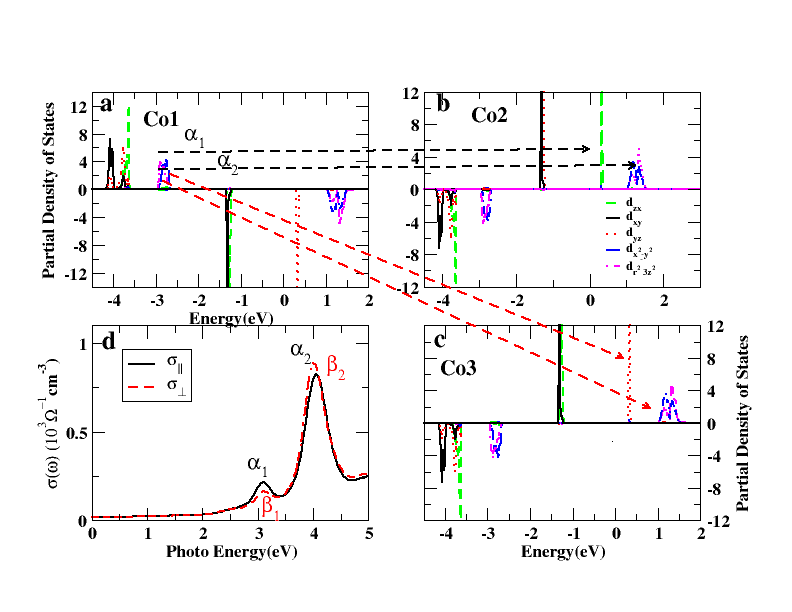}
		\caption{The projected density of states of Co1 (a), Co2 (b), Co3 (c) and the  calculated  optical conductivity (d) of $G$-AF. Where the $\sigma_{\parallel}$ means optical conductivity in $a-b$ plane, while the $\sigma_{\perp}$ means optical conductivity along $c$-axis.
		}
	\end{center}
\end{figure}

\begin{figure}
\centering
\includegraphics[width=1\linewidth]{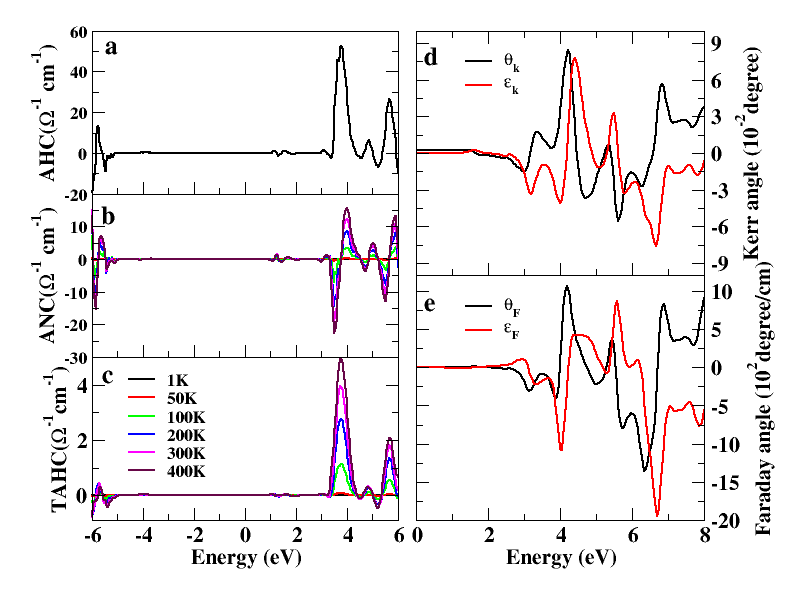}
\caption{\label{fig:ahcanc}
(a) The  anomalous Hall conductivity (AHC). (b) anomalous Nernst conductivity (ANC) and (c) anomalous thermal Hall conductivity (TAHC) varying with  temperature. (d) The magneto-optical Kerr rotation $\theta_k$  and Kerr ellipticity $\epsilon_k$. (e) Faraday rotations  angle $\theta_F$ and Faraday ellipticity $\epsilon_F$. 
}
\end{figure}

In conventional antiferromagnets with preserved $\hat{PT}$ symmetry, the Berry curvature vanishes identically in all directions due to the constraint: $\hat{PT}\Omega_n^{\alpha}(\mathbf{k})  = -\Omega_n^{\alpha}(\mathbf{k}) \implies \Omega_n^{\alpha}(\mathbf{k}) \equiv 0$~\cite{Topandoptical}. In contrast, altermagnet $G$-AFM NaCoF$_3$ breaks the $\hat{PT}$ symmetry while retaining 8 symmetry operations listed in TABLE. II.  $R_{2y}\Omega_n^{y}(\mathbf{k}) = \Omega_n^{y}(-k_x,k_y,-k_z) = +\Omega_n^{y}(\mathbf{k})$, inducing the non-vanished $\Omega_n^{y}(\mathbf{k})$. At the same time,  $\hat{T}R_{2x/z}\Omega_n^{x/z}(\mathbf{k}) = -\Omega_n^{x/z}(\mathbf{k})=0$ results in the trivial   $\Omega_n^{x/z}(\mathbf{k})=0$. For the $G$-AFM NaCoF$_3$,  it only have nonzero $\Omega_n^{y}$  induced AHC, ANC and TAHC as shown in Fig.7.  Around 4 eV, there are sharp peaks in the AHC, ANC and TAHC (Fig.7abc),  indicating the electron-doping  induced anomalous transport properties. At the same time, the Kerr rotation angle  $\theta_k$  and Kerr ellipticity $\epsilon_k$ reach their peaks as high as 9$\times10^{-2}$ degree in Fig.7d, which can be used to  quantitatively verify its  alter-magnetic character by measuring the rotation of the  polarization of reflection light. When probed in transmission geometry, it exhibits an exceptionally large Faraday rotation  up to 1000 degree/cm (Fig.7e).

In conventional antiferromagnets with preserved $\hat{\mathcal{P}\mathcal{T}}$ symmetry, the Berry curvature vanishes identically in all momentum directions due to the constraint $\hat{\mathcal{P}\mathcal{T}}\Omega_n^{\alpha}(\mathbf{k}) = -\Omega_n^{\alpha}(\mathbf{k}) \implies \Omega_n^{\alpha}(\mathbf{k}) \equiv 0$~\cite{Topandoptical}. In contrast, the altermagnetic $G$-type antiferromagnet NaCoF$_3$ breaks $\hat{\mathcal{P}\mathcal{T}}$ symmetry while retaining the eight symmetry operations listed in Table. II, with the residual symmetries enforcing anisotropic Berry curvature components through $\mathcal{R}_{2y}\Omega_n^{y}(\mathbf{k}) = \Omega_n^{y}(-k_x, k_y, -k_z) = +\Omega_n^{y}(\mathbf{k})$ and $\hat{\mathcal{T}}\mathcal{R}_{2x/z}\Omega_n^{x/z}(\mathbf{k}) = -\Omega_n^{x/z}(\mathbf{k}) = 0$, yielding nonvanishing $\Omega_n^{y}(\mathbf{k})$ but vanishing $\Omega_n^{x/z}(\mathbf{k})$. The altermagnetic NaCoF$_3$ exhibits exclusively $y$-direction anomalous transport properties, manifested in three distinct phenomena: the anomalous Hall conductivity (AHC), anomalous Nernst conductivity (ANC), and thermal anomalous Hall conductivity (TAHC). All of them display pronounced peaks centered around 4 eV in their respective spectra as clearly shown in Fig.7(a)-(c).  
The magneto-optical response exhibits distinctive signatures characteristic of alter-magnetism~\cite{yaoMOKE,Topandoptical}, with the Kerr rotation angle $\theta_K$ and ellipticity $\epsilon_K$ attaining maximum values of $9\times10^{-2}$degree and  $8\times10^{-2}$degree(Fig.7d) respectively. While the Kerr rotation angle is smaller than that of ferromagnetic compounds~\cite{Fe2MnSn,CoTiSn,RuF4-Kerr}, approximately 10$\sim$50 times larger than that of $\hat{\mathcal{P}\mathcal{T}}$-symmetry-protected antiferromagnetic materials~\cite{mokeeq1,mokeeq2,mokeeq3,MnBiTe}, and comparable to other altermagnetic compounds~\cite{yaoMOKE,alterKerr1,NANO-Kerr}. These pronounced Kerr effects serve as a quantitative probe of the altermagnetic state through polarization-resolved reflectivity measurements. Furthermore, transmission magneto-optical response reveal an exceptionally large Faraday rotation\cite{Topandoptical} of $1.1\times10^3$ degree/cm (Fig.7e), which exceeds typical values observed in conventional magnetic materials by nearly an order of magnitude~\cite{yaoMOKE,alterKerr1,NANO-Kerr}. The coexistence of these robust magneto-optical responses demonstrates significant potential for applications in altermagnetism-based optical spintronic technologies.

\begin{figure}[htbp]
	\begin{center} 
		\includegraphics[clip,scale=0.4]{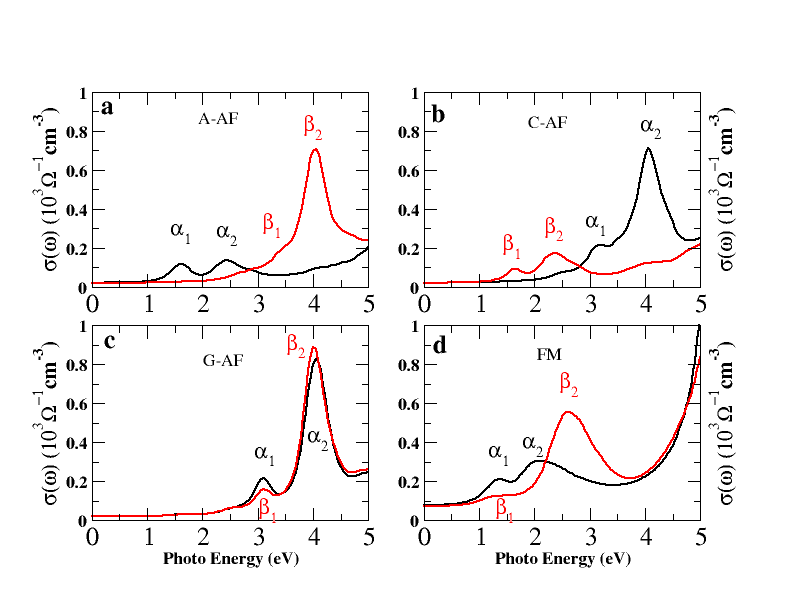}
		\caption{The calculated optical conductivity of  for $A$-AFM phase (a), $C$-AFM phase (b), $G$-AFM phase (c) and FM phase (d). }	
		\end{center}
\end{figure}

Generally, the alter-magnetic character of materials can be conclusively verified through combined measurements of polarized neutron diffraction and angle-resolved photoemission spectroscopy~\cite{Krempasky2024-Nature}. However, we find optical conductivity measurement can also determine the magnetic ordering and the anisotropic electronic structure. In Fig.8, we show the calculated optical conductivity of different magnetic phases. In the $A$-AFM (Fig.8(a)) and FM (Fig.8(d)) phases, where the ferromagnetic ordering preserves in a-b plane, which provides the passageway for the electron hopping
Co$_{1}$-d$^{\uparrow}_{xy/zx}\leftrightarrow$ Co$_{2}$-d$^{\uparrow}_{zx}$ and Co$_{1}$-d$^{\uparrow}_{{x^{2}-y^{2}/{r^{2}-3z^{2}}}}\leftrightarrow$ Co$_{2}$-d$^{\uparrow}_{zx}$. So the peak $\alpha_{1}$ and $\alpha_{2}$ appears in both phases. For the same reason, $\beta_{1}$ and $\beta_{2}$ appear in the $C$-AFM, because of the parallel magnetic moments along $c$-axis. The magneto-optical conductivity spectra of other magnetic states can be well understood through the combination of three distinct types of magnetic ordering and orbital ordering (see Supplementary Material).

\subsection{Summary and Conclusion}
Combining first-principles calculations with systematic symmetry analysis, we investigate the electronic, magnetic, and orbital structures of altermagnetic perovskite compounds, using NaCoF$_3$ as a prototypical example. Our GGA+U calculations establish a $G$-type antiferromagnetic ground state with concomitant $G$-type orbital ordering. The interplay between symmetry-protected momentum-space spin splitting and real-space anisotropic charge distribution generates characteristic altermagnetic responses, including the anomalous Hall effect (AHE), anomalous Nernst effect (ANE), anomalous thermal Hall effect (ATHE), and significant magneto-optical (Kerr/Faraday) responses. Fermi surface analysis reveals $d$-wave topology with pronounced spin-splitting away from high-symmetry planes. Optical conductivity calculations identify distinct inter-orbital transitions between Co$^{2+}$ sites, while comparative studies across magnetic phases demonstrate that optical conductivity measurements provide a sensitive experimental probe for determining magnetic ordering configurations.

\begin{acknowledgments}
The authors  acknowledge the supports from NSF of China (No.11904084 and No.10947001) and the Innovation Scientists and Technicians Troop Constriction Projects of Henan Province (Grant No. 104200510014).
\end{acknowledgments}

\bibliography{reference}

\newpage

{
\centering
\LARGE
\textbf{SUPPLEMENTARY MATERIALS}
\par
}
\vspace{1em} 

In the Supplementary Materials, we provide detailed information on two additional antiferromagnetic configurations (A-AFM and C-AFM), including symmetry analysis, topological features of the Fermi surface, spin-induced band splitting, and calculations of the Faraday and Kerr effects. Also included are the crystal structures and electronic band structures for eight other collinear antiferromagnetic materials.


\section{Results of A-AFM}

For the \(A\)-AFM configuration, its symmetry operations can be divided into two categories. The first group consists of:

\[
\{ E, I, \left\{ R_{2x}\mid\tfrac{1}{2}\tfrac{1}{2}0\right\}, \left\{ M_x\mid\tfrac{1}{2}\tfrac{1}{2}0 \right\} \}
\]

The second group consists of:

\[
\left\{ \hat{T} \left\{ R_{2y}\mid\tfrac{1}{2}\tfrac{1}{2}\tfrac{1}{2}\right\}\hat{T}\left\{ M_y \mid \tfrac{1}{2}\tfrac{1}{2}\tfrac{1}{2}\right\}, \hat{T}\left\{ R_{2z}\mid00\tfrac{1}{2}\right\}, \hat{T} \left\{ M_z \mid 00\tfrac{1}{2} \right\} \right\}
\]

The first group corresponds to transformations that interchange atoms within a single spin sub-lattice, while the second group symmetries interchange atoms between opposite-spin sub-lattices and ensure zero net magnetization.

\[
\hat{T} \left[ R_{2y} \mid \tfrac{1}{2} \tfrac{1}{2} \tfrac{1}{2} \right] E(k_x, k_y, k_z, s) =E(k_x, -k_y, k_z, -s)
 \mathop{=}_{k_y = \pm \frac{\pi}{2}}^{k_y = 0}E(k_x, k_y, k_z, -s) 
\]

\[
\hat{T} \left[ R_{2z} \mid 00 \tfrac{1}{2} \right] E(k_x, k_y, k_z, s) =E(k_x, k_y, -k_z, -s)
 \mathop{=}_{k_z = \pm \frac{\pi}{2}}^{k_z = 0}E(k_x, k_y, k_z, -s) 
\]
So, at the $k_y$=0,$\pm \frac{\pi}{2}$ and $k_z$=0,$\pm \frac{\pi}{2}$ planes, the spin-up and spin-down bands are degenerated, as shown in the Fig.S1a/c. At the $k_x$=0,$\pm \frac{\pi}{2}$ planes, no symmetry operation protecting the spin degeneracy. When the  $\hat{T}\{R_{2z}|00\frac{1}{2}\}$ symmetry is broken due to deviation from the $k_z$=0($\frac{\pi}{2}$) mirror plane, the spin degeneracy is lifted, leading to characteristic band splitting as illustrated in Fig.S1d.  To explicitly address the band splitting phenomenon, we plot the band structure along the  AZ$\bar{A}$-line in  Fig.S2a and its  magnified views around the band energy 1.38 eV  in Fig. S2(b) and -1.2 eV  in Fig. S2(c). 

\renewcommand{\thefigure}{S\arabic{figure}} 
\setcounter{figure}{0}
\begin{figure}[!htb]
	\begin{center}	
		\includegraphics[clip,scale=0.4]{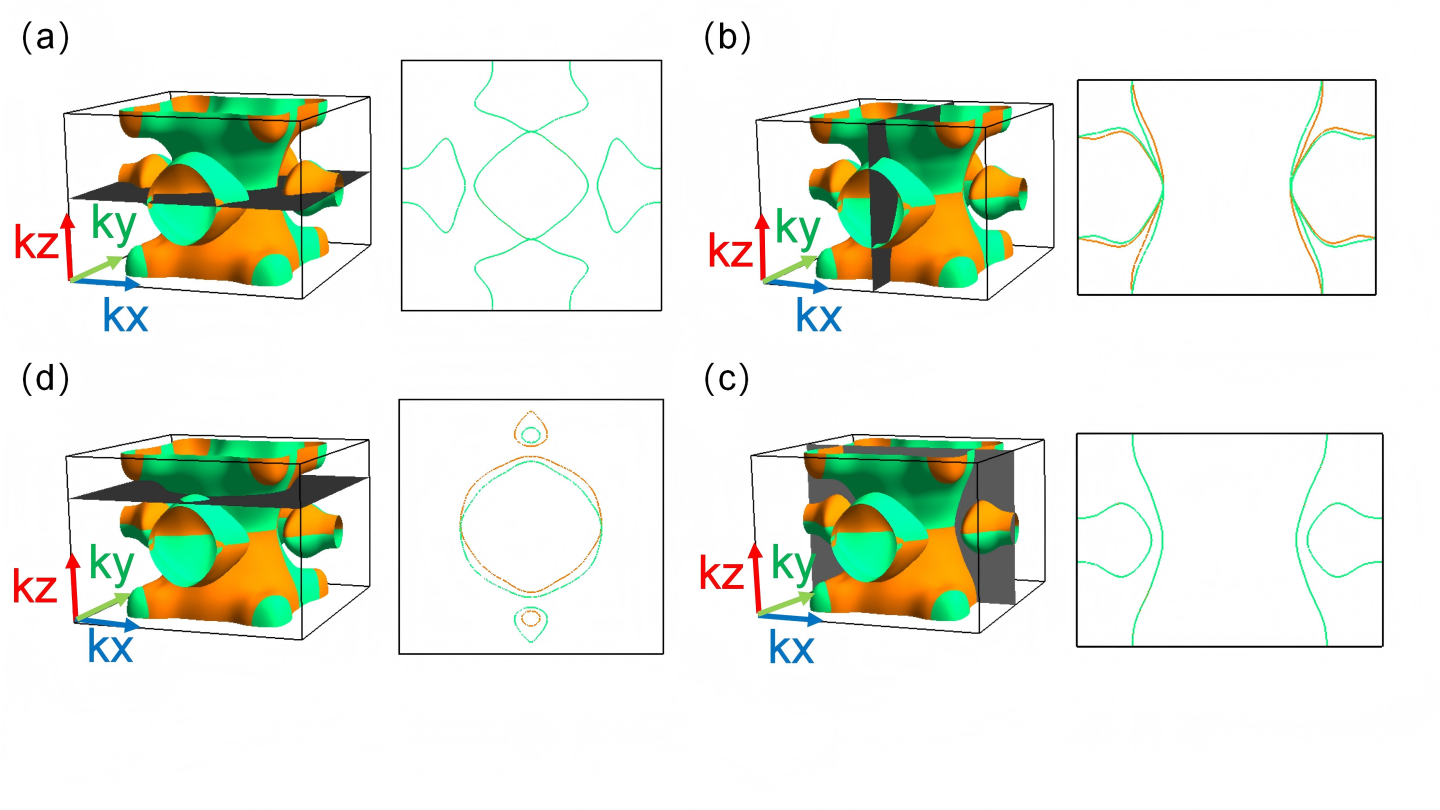}
		\caption{For the A-AFM, (a), (b)  and (c) are the Fermi surfaces (set Fermi level to -1.2eV) and their cut-sections at  the $k_z$=0, $k_x$=0 and $k_y$=0 plane.  (d)  is  the surface and its cut-section at  the $k_z$=0.25. 
	}
	\end{center}
\end{figure}

%

\begin{figure}[!htb]
	\begin{center} 
 	\includegraphics[clip,scale=0.4]{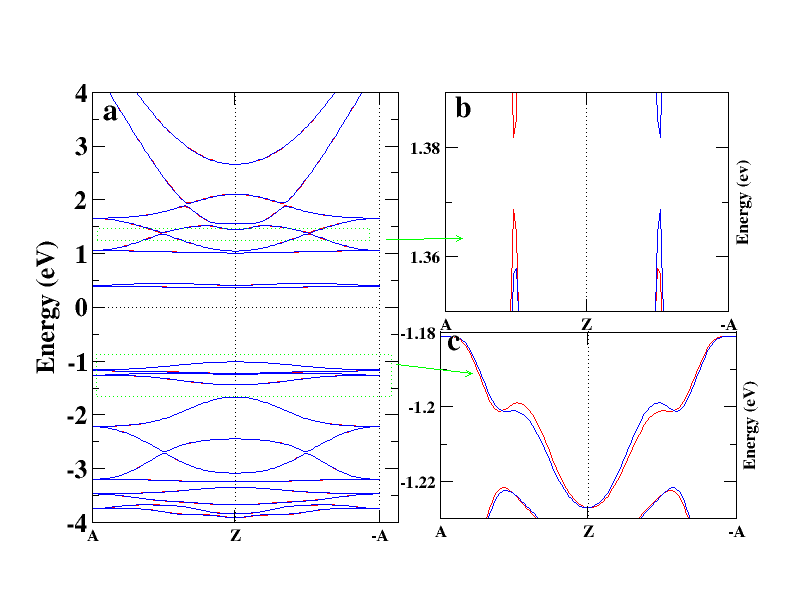}
		\caption{ For the A-AFM, figure (a) displays the band structure calculated along the general $k$-path in the  $k_z$=0.35 plane, where the spin-up and spin-down bands are not degenerated. The splitting of the band around 1.35 eV and -1.28 eV are presented in figure (b) and (c).
		}
	\end{center}
\end{figure}

Spin-resolved charge density in Fig.S3 reveals the magnetic and orbital order in the   $A$-AFM NaCoF$_3$. In the $a-b$ plane,  the  nearest-neighbor Co$^{2+}$ ions exhibit parallel spin alignment  coupled with antiparallel orbital ordering.  Along $c$-axis, both magnetic coupling and orbital order are anti-parallel.

Symmetry analysis demonstrates that Co$_1$ and Co$_3$ are mutually mapped through  $\hat{T}\{R_{2z}|00\frac{1}{2}\}$  symmetry operation, while Co$_1$ and Co$_2$ are correlated via $\hat{T}\{R_{2y}|\frac{1}{2}\frac{1}{2}\frac{1}{2}\}$  symmetry operation.

\begin{figure}[!htb]
	\begin{center} 
		\includegraphics[clip,scale=0.2]{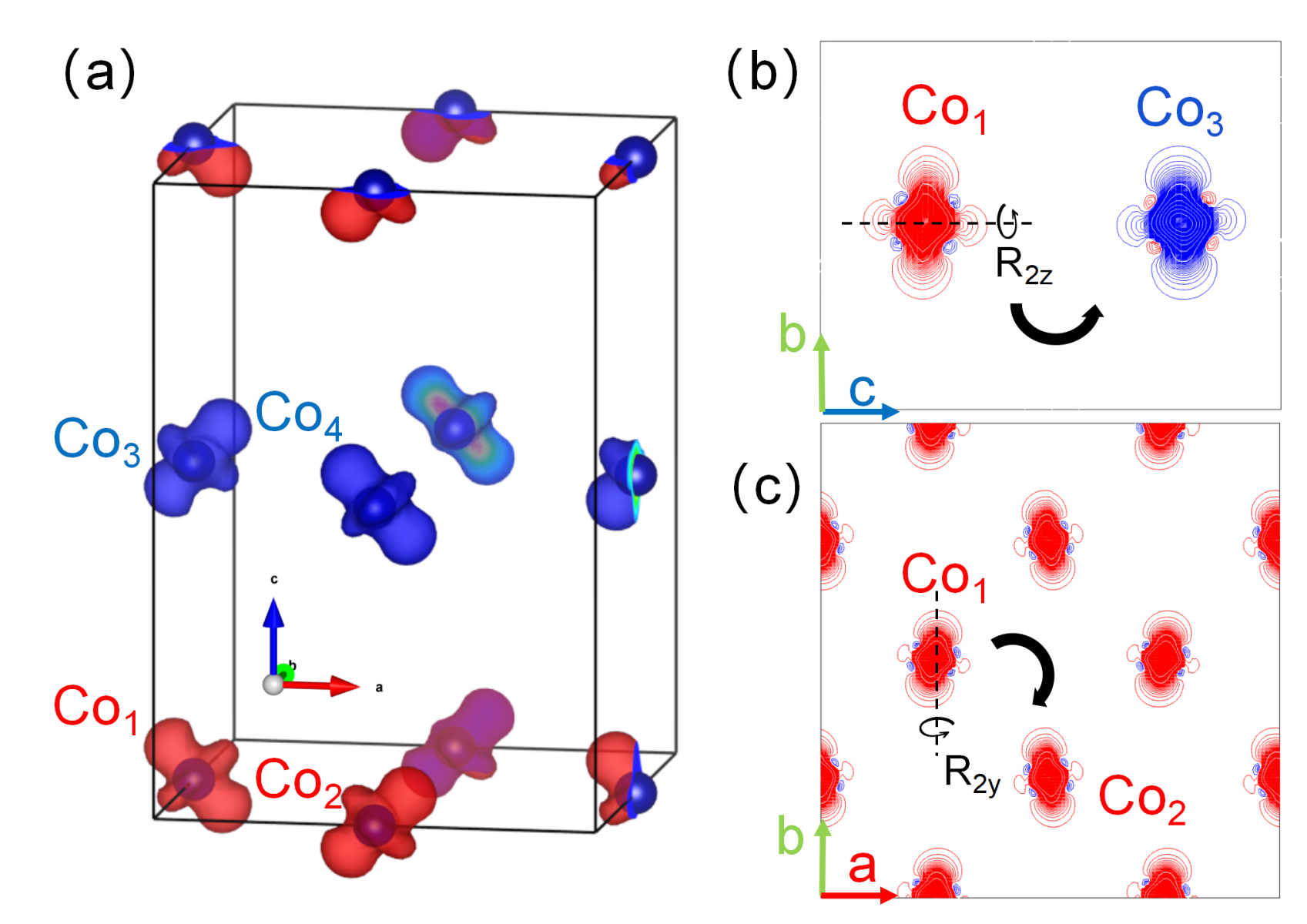}
		\caption{ The charge density of $A$-AFM state, with their energy around 0.35 eV (the flat band). (a) is the 3D charge density, while (b) and (c) are 2D charge density on the 100-plane and 001-plane, respectively.   Red  for up-spin, blue for down-spin.
		}	
	\end{center}
\end{figure}

\begin{figure}[!htb]
	\begin{center} 
		\includegraphics[clip,scale=0.4]{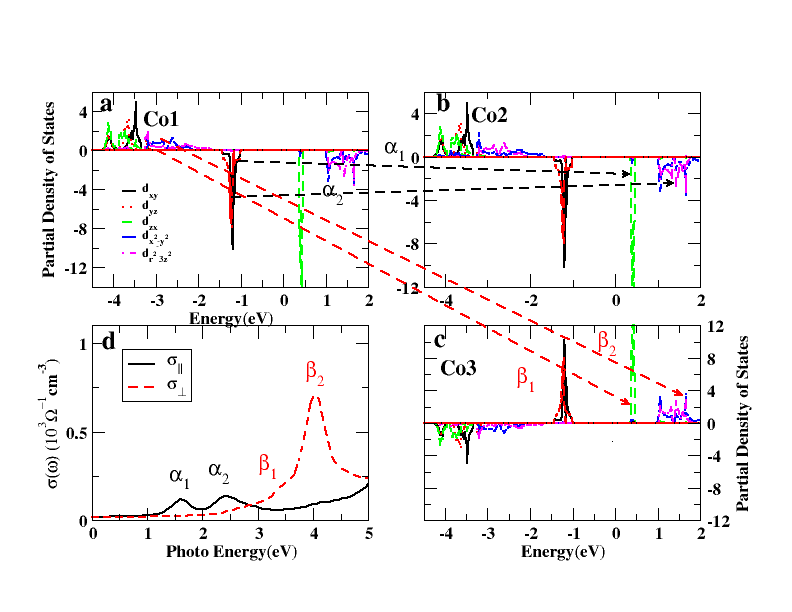}
		\caption{The projected density of states of Co1 (a), Co2 (b), Co3 (c) and the  calculated  optical conductivity (d) of \(A\)-AFM. Where the $\sigma_{\parallel}$ means optical conductivity in a-b plane, while the $\sigma_{\perp}$ means optical conductivity along $c$-axis.
		}
	\end{center}
\end{figure}

The  anisotropic Fermi surface and spin-split band structure in the \(A\)-AFM state has been explained by the symmetry operations, charge  and orbital order. But such properties are difficult to be measured. The  anisotropic optical conductivity measurement is easy to be performed, and can be understood by the projected density of states   of three distinct Co$^{2+}$ ions. Compared with the \(G\)-AFM state, the difference in spin orientation induces significant shifts in photoconductivity peak positions: within the a-b plane, two characteristic peaks are observed - the $\alpha_{1}$ peak at 1.6 eV and $\alpha_{2}$ peak at 2.4 eV – corresponding to transitions from occupied Co$_1$-d$^{\downarrow}_{xy/yz}$ to unoccupied Co$_2$-d$^{\downarrow}_{zx}$, and from e$^{\downarrow}_{g}$dorbitals respectively ( black-dashed arrows). Along the c-axis, the $\alpha_{1}$ and $\alpha_{2}$ peaks vanish due to spin-flip selection rules, while the emergence of new characteristic peaks at 3.5 eV and 4.0 eV, designated as $\beta_{1}$ and $\beta_{2}$ peaks respectively, originates from distinct inter-ionic electronic transitions: the $\beta_{1}$ peak arises from Co$_1$-e$^{\uparrow}_{g}$ to Co$_3$-d$^{\uparrow}_{zx}$ transitions, while the peak $\beta_{2}$ corresponds to transitions between  Co$_1$-e$^{\uparrow}_{g}$ and Co$_3$-e$^{\uparrow}_{g}$ (red-dashed arrows).

\section{Results of C-AFM}

For the \(C\)-AFM configuration, its symmetry operations can be divided into two categories. The first group consists of:

\[
\left\{
E,\ 
I,\ 
\left\{R_{2z}\mid00\tfrac{1}{2}\right\},\ 
\left\{M_z\mid00\tfrac{1}{2}\right\}
\right\}
\]

The second group consists of:

\[
\left\{
\hat{T}\left\{R_{2x}\mid\tfrac{1}{2}\tfrac{1}{2}0\right\},\ 
\hat{T}\left\{R_{2y}\mid\tfrac{1}{2}\tfrac{1}{2}\tfrac{1}{2}\right\},\ 
\hat{T}\left\{M_x\mid\tfrac{1}{2}\tfrac{1}{2}0\right\},\ 
\hat{T}\left\{M_y\mid\tfrac{1}{2}\tfrac{1}{2}\tfrac{1}{2}\right\}
\right\}
\]

The first group corresponds to transformations that interchange atoms within a single spin sub-lattice, while the second group symmetries interchange atoms between opposite-spin sub-lattices and ensure zero net magnetization.

\[
\hat{T} \left[ R_{2x} \mid \tfrac{1}{2} \tfrac{1}{2}0 \right] E(k_x, k_y, k_z, s) =E(-k_x, k_y, k_z, -s)
 \mathop{=}_{k_x = \pm \frac{\pi}{2}}^{k_x = 0}E(k_x, k_y, k_z, -s) 
\]

\[
\hat{T} \left[ R_{2y} \mid \tfrac{1}{2}\tfrac{1}{2}\tfrac{1}{2}\right] E(k_x, k_y, k_z, s) =E(k_x, -k_y, k_z, -s)
 \mathop{=}_{k_y = \pm \frac{\pi}{2}}^{k_y = 0}E(k_x, k_y, k_z, -s) 
\]
\begin{figure}[!htb]
	\begin{center}
	\includegraphics[clip,scale=0.4]{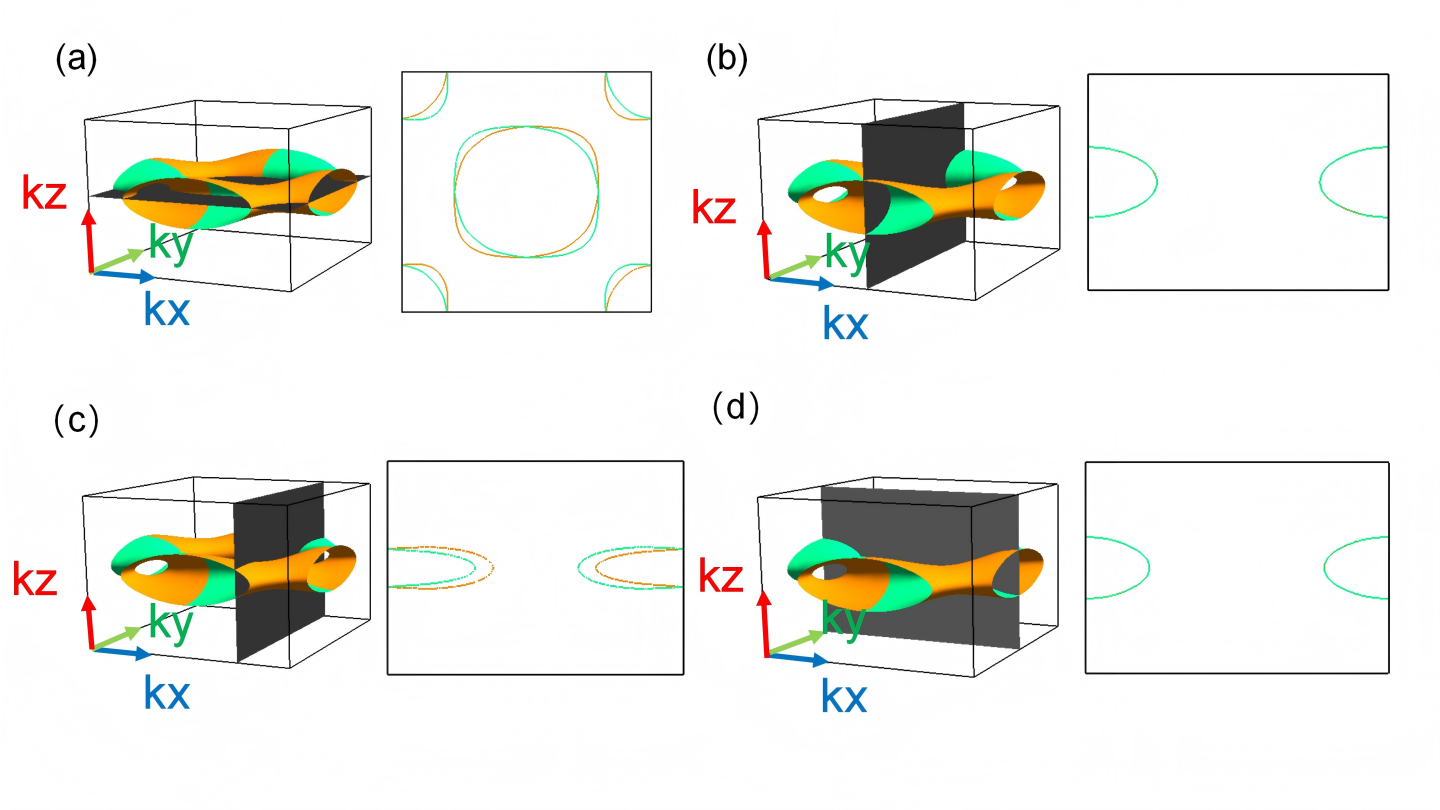}
		\caption{ For the C-AFM, (a), (b)  and (c) are the Fermi surfaces (set Fermi level to -1.2eV) and their cut-sections at  the $k_z$=0, $k_x$=0 and $k_y$=0 plane.  (d)  is  the surface and its cut-section at  the $k_x$=0.25. 
				}
	\end{center}
\end{figure}

\begin{figure}[!htb]
	\begin{center} 
	\includegraphics[clip,scale=0.25]{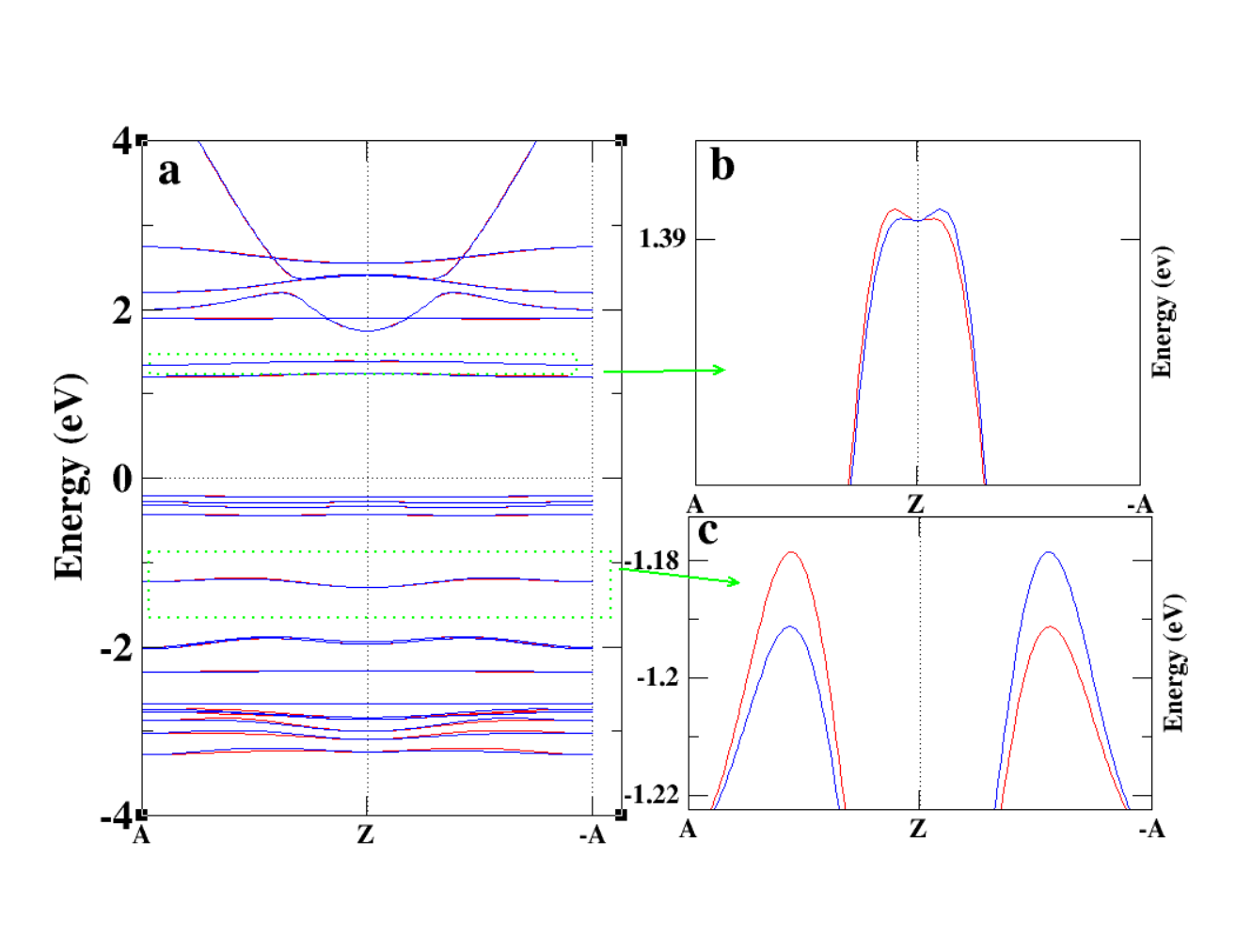}
		\caption{  For the C-AFM, figure (a) displays the band structure calculated along the general $k$-path in the  $k_z$=0 plane, where the spin-up and spin-down bands are not degenerated. The splitting of the band around 1.39 eV and -1.2 eV are presented in figure (b) and (c).
				}
	\end{center}
 \end{figure}

\begin{figure}[!htb]
	\begin{center} 
		\includegraphics[clip,scale=0.15]{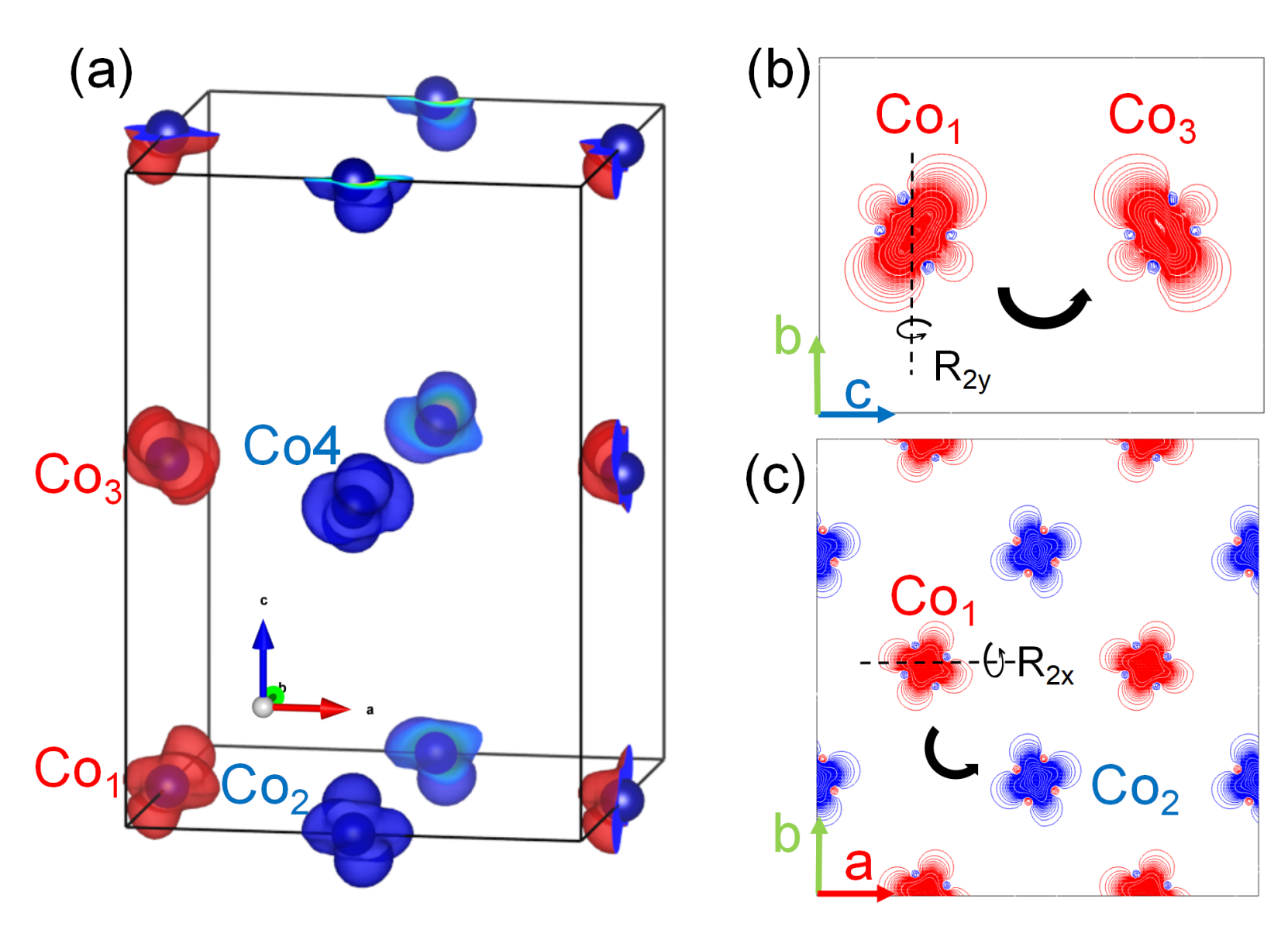}
		\caption{ The charge density of $C$-AFM state, with their energy around 0.35 eV (the flat band just above the Fermi energy). (a) is the 3D charge density, while (b) and (c) are 2D charge density on the 100-plane and 001-plane, respectively.    Red  for up-spin, blue for down-spin.
		\label{fig:S7}}	
	\end{center}
\end{figure}

At the \( k_x = 0, \pm \frac{\pi}{2} \) and \( k_y = 0, \pm \frac{\pi}{2} \) planes, the spin-up and spin-down bands are degenerate, as shown in Figs. S5b-c. In contrast, at the \( k_z = 0, \pm \frac{\pi}{2} \) planes, the Fermi surfaces are split due to the lack of a symmetry operation protecting the spin degeneracy. When the \( \hat{T}\left\{R_{2x} \mid \frac{1}{2}\frac{1}{2}0\right\} \) symmetry is broken, caused by a deviation from the \( k_x = 0 \) (\( \frac{\pi}{2} \)) mirror plane, spin degeneracy is lifted, resulting in the split Fermi surfaces shown in Fig. S5c and the characteristic band splitting illustrated in Fig. S6. To further explore this band splitting, we plot the band structure along the AZ\(\bar{A}\)-line in Fig. S6a, along with magnified views around the characteristic band energies of 1.39 eV (Fig. S6b) and -1.2 eV (Fig. S6c).

To elucidate the electronic structure characteristics of the $C$-AFM state, we present its spin-resolved charge density distribution in Fig.~\ref{fig:S7}. Unlike the $A$-AFM state, the $C$-AFM NaCoF$_3$ exhibits a distinctive configuration, where nearest-neighbor Co$^{2+}$ ions show antiferromagnetic spin alignment and antiparallel orbital ordering in the a-b plane. Along the c-axis, the magnetic order is ferromagnetic with antiparallel orbital alignment. Symmetry analysis reveals that Co$_1$ and Co$_3$ are interconvertible through the symmetry operations \( \hat{T}\{R_{2y}\mid\frac{1}{2}\frac{1}{2}\frac{1}{2}\} \) or \( \hat{T}\{M_y\mid\frac{1}{2}\frac{1}{2}\frac{1}{2}\} \), while Co$_1$ and Co$_2$ are correlated via \(\hat{T}\{R_{2x}\mid \frac{1}{2}\frac{1}{2}0\} \) or \( \hat{T}\{M_x\mid \frac{1}{2}\frac{1}{2}0\} \).

\begin{figure}[!htb]
	\begin{center} 
		\includegraphics[clip,scale=0.4]{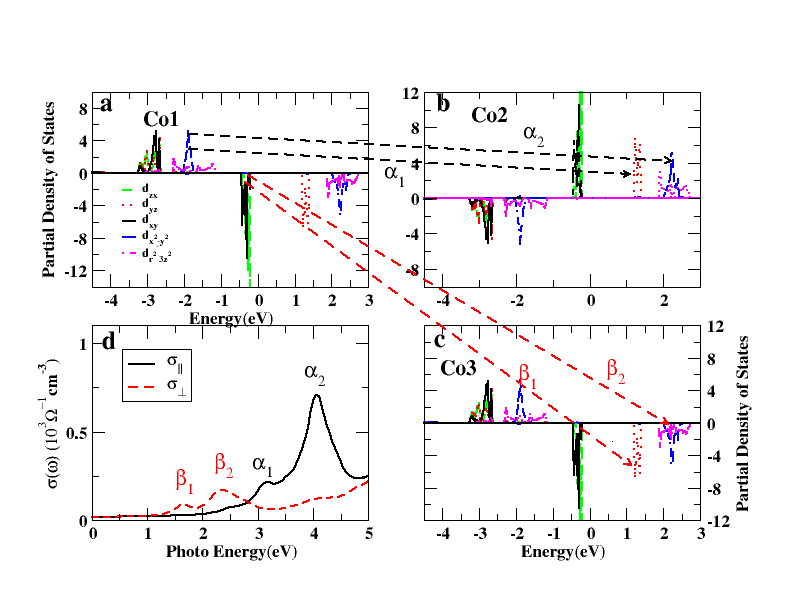}
		\caption{The projected density of states for Co$_1$ (a), Co$_2$ (b), Co$_3$ (c), and the calculated optical conductivity (d) of the $C$-AFM state are shown. Here, \( \sigma_{\parallel} \) represents the optical conductivity in the a-b plane, while \( \sigma_{\perp} \) denotes the optical conductivity along the $c$-axis.
		}
	\end{center}
\end{figure}

As shown in Fig. S8, the projected density of states and optical conductivity analysis of three Co$^{2+}$ ions reveal the distinctive electronic structure characteristics of the $C$-AFM state. In the $ab$-plane, the characteristic peaks at 3.2 eV (denoted as $\alpha_{1}$) and 4.1 eV (denoted as $\alpha_{2}$) originate from transitions between the occupied Co$_1$-e$^{\uparrow}_{g}$ and the unoccupied Co$_2$-d$^{\uparrow}_{yz}$ and Co$_2$-e$^{\uparrow}_{g}$ orbitals, as indicated by the black arrows. Along the $c$-axis, the peaks at 1.7 eV (denoted as $\beta_{1}$) and 2.4 eV (denoted as $\beta_{2}$) correspond to transitions from Co$_1$-d$^{\downarrow}_{zx/xy}$ to Co$_3$-d$^{\downarrow}_{yz}$ and Co$_3$-e$^{\downarrow}_{g}$, marked by red arrows. These spectral features show notable differences in both energy positions and orbital origins compared to the $A$-AFM and $G$-AFM states, highlighting the unique electronic structure under different magnetic orderings.

\section{ the magneto-optic Kerr  and Faraday effect}

In studies of the magneto-optic Kerr  and Faraday effect, the Kerr rotation angle $\theta_k$ and Kerr ellipticity $\epsilon_k$ are typically combined into a complex-valued quantity known as the complex Kerr angle, expressed as:
\begin{eqnarray}
\phi^{\gamma}_{K} = \vartheta^{\gamma}_{K} + i\varepsilon^{\gamma}_{K} \approx \frac{-\nu_{\alpha\beta\gamma} \sigma_{\alpha\beta}}{\sigma_{0} \sqrt{1 + i(4\pi/\omega)\sigma_{0}}} ,  
 \end{eqnarray}
Here, $\alpha$ and $\beta$ ,$\gamma$(=$x,y,z$), $\nu_{\alpha\beta\gamma}$ is the Levi-Civita symbol with the Cartesian coordinates. $\sigma_0 = (\sigma_{\alpha\alpha} +\sigma_{\beta\beta})/2$ represents the average diagonal component of the conductivity tensor, typically approximated by $\sigma_{\alpha\alpha}$. 
The complex Faraday angle can be expressed as:
\begin{eqnarray}
 \phi^{\gamma}_{F} = \vartheta^{\gamma}_{F} + i\varepsilon^{\gamma}_{K} \approx\nu_{\alpha\beta\gamma} \frac{(n_+ - n_-)\omega d}{2c} , 
 \end{eqnarray}
 For bulk materials, the thickness $d$ is not required to be specified since the Faraday angles are calculated in units of deg/cm. The complex refractive indices $n_\pm$ are expressed as:
 \begin{equation}
 	n_\pm = \sqrt{1 + \frac{4\pi i}{\omega} (\sigma_{\alpha\alpha} \pm i \sigma_{\alpha\beta})},
 \end{equation}
 where $\sigma_{\alpha\alpha}$ and $\sigma_{\alpha\beta}$ are the diagonal and off-diagonal components of the conductivity tensor, respectively. Under the condition that $\sigma_{\alpha\alpha} \gg \sigma_{\alpha\beta}$, the complex refractive indices $n_\pm$ can be approximated as:
 \begin{equation}
 	n_\pm \approx \left[1 + \frac{4\pi i}{\omega} \sigma_{\alpha\alpha}\right]^{1/2} \mp \frac{2\pi}{\omega} \sigma_{\alpha\beta} \left[1 + \frac{4\pi i}{\omega} \sigma_{\alpha\alpha}\right]^{-1/2}.
 \end{equation}
 Therefore, the complex Faraday angle $\theta_F$ can be approximated as:
 \begin{eqnarray}
 \phi^{\gamma}_{F} = \vartheta^{\gamma}_{F} + i\varepsilon^{\gamma}_{K} = \frac{-\nu_{\alpha\beta\gamma} \sigma_{\alpha\beta}}{ \sqrt{1 + i(4\pi/\omega)\sigma_{0}}} \frac{2\pi d}{c} , 
 \end{eqnarray}

 In conventional antiferromagnetic materials, the Berry curvature vanishes in all momentum directions due to the preservation of \( \hat{PT} \) symmetry. However, in the $G$-type antiferromagnetic altermagnetic NaCoF$_3$, the system breaks \( \hat{PT} \) symmetry while retaining the symmetry operations \( \{R_{2y} \mid \frac{1}{2}\frac{1}{2}\frac{1}{2}\} \), \( \hat{T} \{R_{2x} \mid \frac{1}{2}\frac{1}{2}0\} \), and \( \hat{T} \{R_{2z} \mid 00\frac{1}{2}\} \). This leads to a strictly anisotropic distribution of Berry curvature: \( \Omega_n^{x/z} = 0 \), with nonzero \( \Omega_n^{y} \) only. This anisotropy significantly enhances the magneto-optical response components of the complex Kerr and Faraday angles along the $y$-direction.

In $A$-AFM systems, the symmetry-breaking pattern preserves the operations \( \{R_{2x}\mid\frac{1}{2}\frac{1}{2}0\} \), \( \hat{T}\{R_{2y}\mid\frac{1}{2}\frac{1}{2}\frac{1}{2}\}\), and \( \hat{T}\{R_{2z}\mid 00\frac{1}{2}\} \). From the symmetry constraints \( R_{2x}\Omega_n^{x}(\mathbf{k}) = \Omega_n^{x}(k_x, -k_y, -k_z) = +\Omega_n^{x}(\mathbf{k}) \neq 0 \) and \( \hat{T} R_{2y/z} \Omega_n^{y/z}(\mathbf{k}) = -\Omega_n^{y/z}(\mathbf{k}) = 0 \), our calculations confirm that only the \( \Omega_n^{x} \) component is nonzero.

 \begin{figure}[!htb]
 	\begin{center} 
 		\includegraphics[clip,scale=0.4]{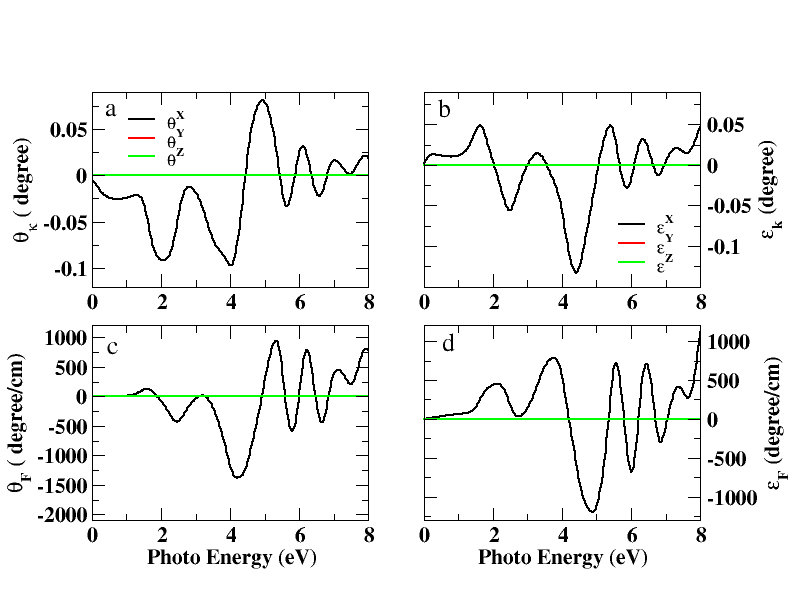}
 		\caption{For the $A$-AFM NaCoF$_3$ (a) The magneto-optical Kerr rotation $\theta_k$ . (b) Kerr ellipticity $\epsilon_k$. (c) Faraday rotations angle $\theta_F$ .(d) Faraday ellipticity $\epsilon_F$ . 
 		}
 	\end{center}
 \end{figure}

As shown in Fig. S9, the $A$-AFM NaCoF$_3$  exhibits a prominent peak in the Kerr rotation angle \( \theta_k \) at 5 eV, reaching \( 8 \times 10^{-2} \) degrees, while the maximum Kerr ellipticity \( \epsilon_k \) of \( 4.8 \times 10^{-2} \) degrees occurs at 1.6 eV. Precise measurements of the reflected light's polarization rotation allow for quantitative characterization of its altermagnetic properties. Notably, in transmission geometry, the material demonstrates an exceptionally large Faraday rotation effect, with a magnitude of 1000 degrees/cm.

\begin{figure}[!htb]
	\begin{center} 
		\includegraphics[clip,scale=0.4]{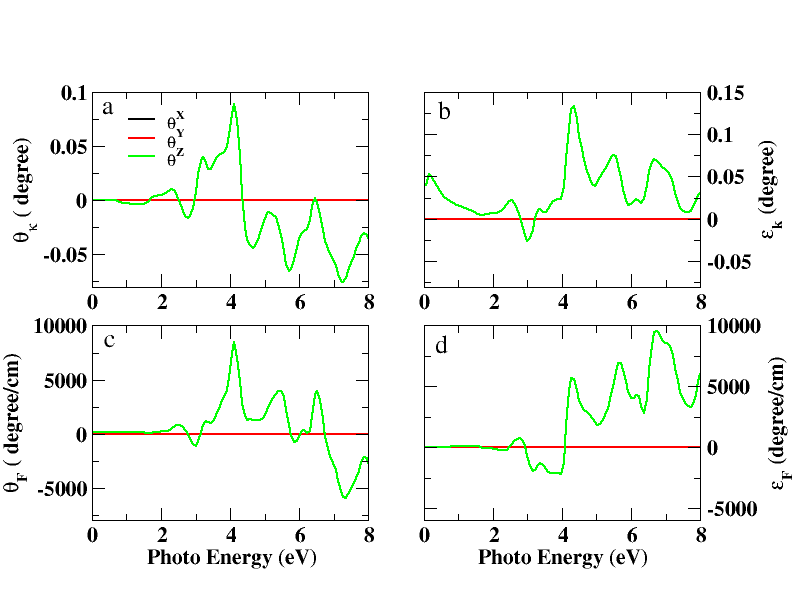}
		\caption{For the $C$-AFM NaCoF$_3$, (a) The magneto-optical Kerr rotation $\theta_k$ . (b) Kerr ellipticity $\epsilon_k$. (c) Faraday rotations angle $\theta_F$ .(d) Faraday ellipticity $\epsilon_F$ . 
		}
	\end{center}
\end{figure}

In $C$-type antiferromagnetic ($C$-AFM) systems, the breaking of $\hat{PT}$ symmetry while preserving the symmetries$\{R_{2z}|00\frac{1}{2}\}$, $\hat{T}\{R_{2y}|\frac{1}{2}\frac{1}{2}\frac{1}{2}\}$and $\hat{T}\{R_{2x}|\frac{1}{2}\frac{1}{2}0\}$ results in a selective distribution of the Berry curvature: under the operation $\{R_{2z}|00\frac{1}{2}\}$, it satisfies $R_{2z}\Omega_n^{z}(\mathbf{k}) = \Omega_n^{z}(-k_x,-k_y,k_z) = +\Omega_n^{z}(\mathbf{k}) \neq 0$, whereas under the operations$\hat{T}R_{2x/y}$, it obeys $\hat{T}R_{2x/y}\Omega_n^{x/y}(\mathbf{k}) = -\Omega_n^{x/y}(\mathbf{k})=0$.

As shown in Figure S10,  the $C$-AFM NaCoF$_3$ exhibits a prominent peak in the Kerr rotation angle $\theta_k$ at 4.1 eV, reaching 9$\times10^{-2}$ degrees, while a maximum Kerr ellipticity $\epsilon_k$ of 1.3$\times10^{-}$ degrees is observed at 4.3 eV. Additionally, the material demonstrates an exceptionally large Faraday rotation effect with a magnitude of 9000 degrees/cm.

In Table.I, we present the total energies, magnetic configurations, and band structures of several perovskite compounds (SrOsO$_3$, SrRhO$_3$, SrRuO$_3$, MnNCl$_3$, NaOsO$_3$, MnTeO$_3$, MnSeO$_3$, and CrBiO$_3$) crystallizing in the space group \textit{Pnma}. SrOsO$_3$ adopts the $A$-AFM ground state, while SrRhO$_3$ and SrRuO$_3$ favor the FM ground state. The remaining compounds exhibit a $G$-AFM ground state. Notably, $C$-AFM SrRhO$_3$ displays the largest spin splitting, reaching 0.3 eV.

\onecolumngrid
	\renewcommand{\arraystretch}{1.4} 
	\setlength{\tabcolsep}{8pt}      
	\begin{longtable}{cccc}
		\caption{The total energy, magnetic structure and band structure of other Perovskite compounds (SrOsO$_3$,  SrRhO$_3$ ,  SrRuO$_3$ , MnNCl$_3$,  NaOsO$_3$,     MnTeO$_3$,   MnSeO$_3$, CrBiO$_3$  ) in the space group  \textit{Pnma}.}  \\
		\toprule
		\textbf{Compound} & \textbf{Ground State ($U=3$)} & \textbf{Altermagnetic Structure} & \textbf{Band} \\
   		\hline
		\endfirsthead
		\caption{The magnetic structure and band characteristics comparison of materials (Continued)} \\
		\toprule
		\textbf{Compound} & \textbf{Ground State ($U=3$)} & \textbf{Altermagnetic Structure} & \textbf{Band} \\
		\hline
		\endhead
        SrOsO$_3$ & 
		\begin{tabular}{@{}c@{}}
			Antiferromagnetism \\
			Energy (meV): \\
			A=0 \\ C=403.72 \\
			G=21.94 \\ FM=62.9
		\end{tabular} & 
		\raisebox{-.6\height}{\includegraphics[height=4cm,width=3.5cm,keepaspectratio]{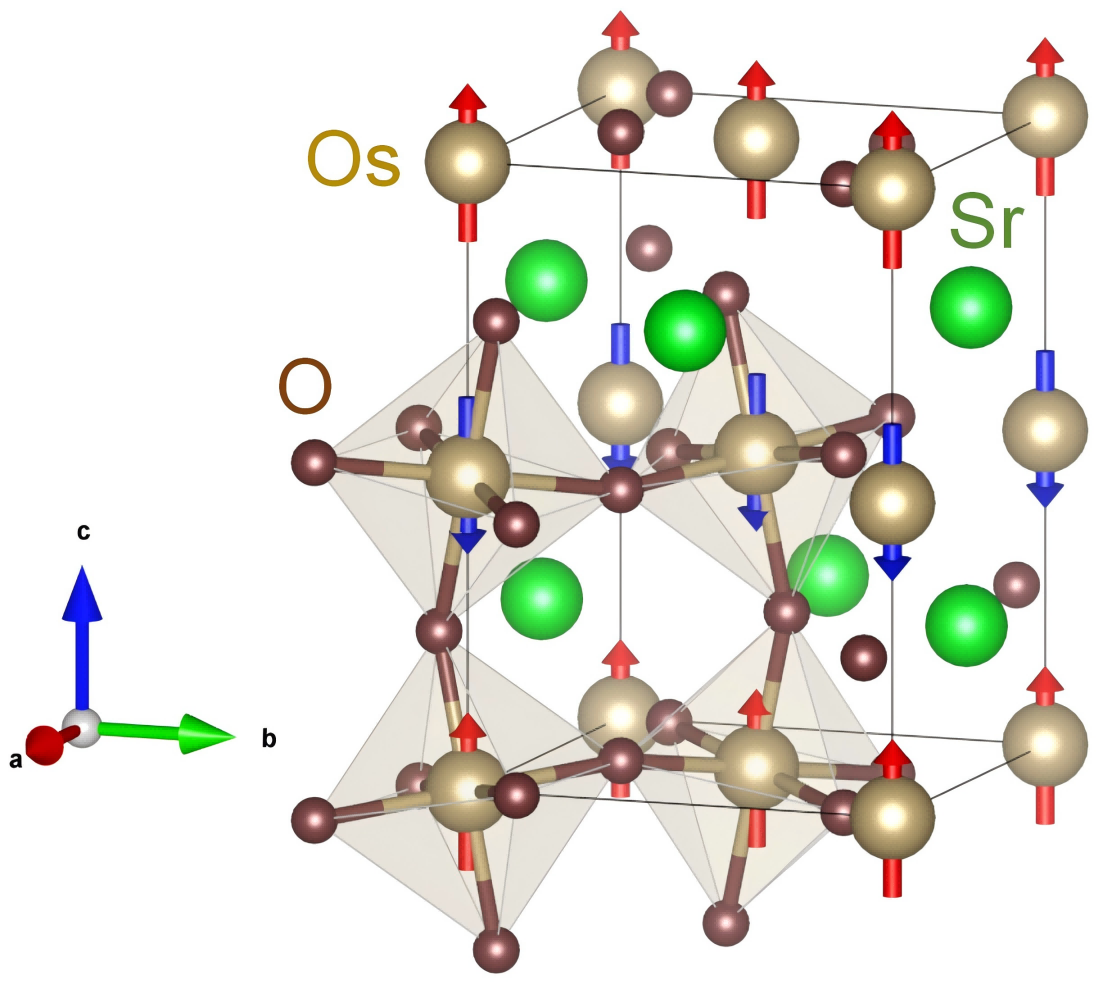}} & 
		\raisebox{-.6\height}{\includegraphics[height=4cm,width=3.5cm,keepaspectratio]{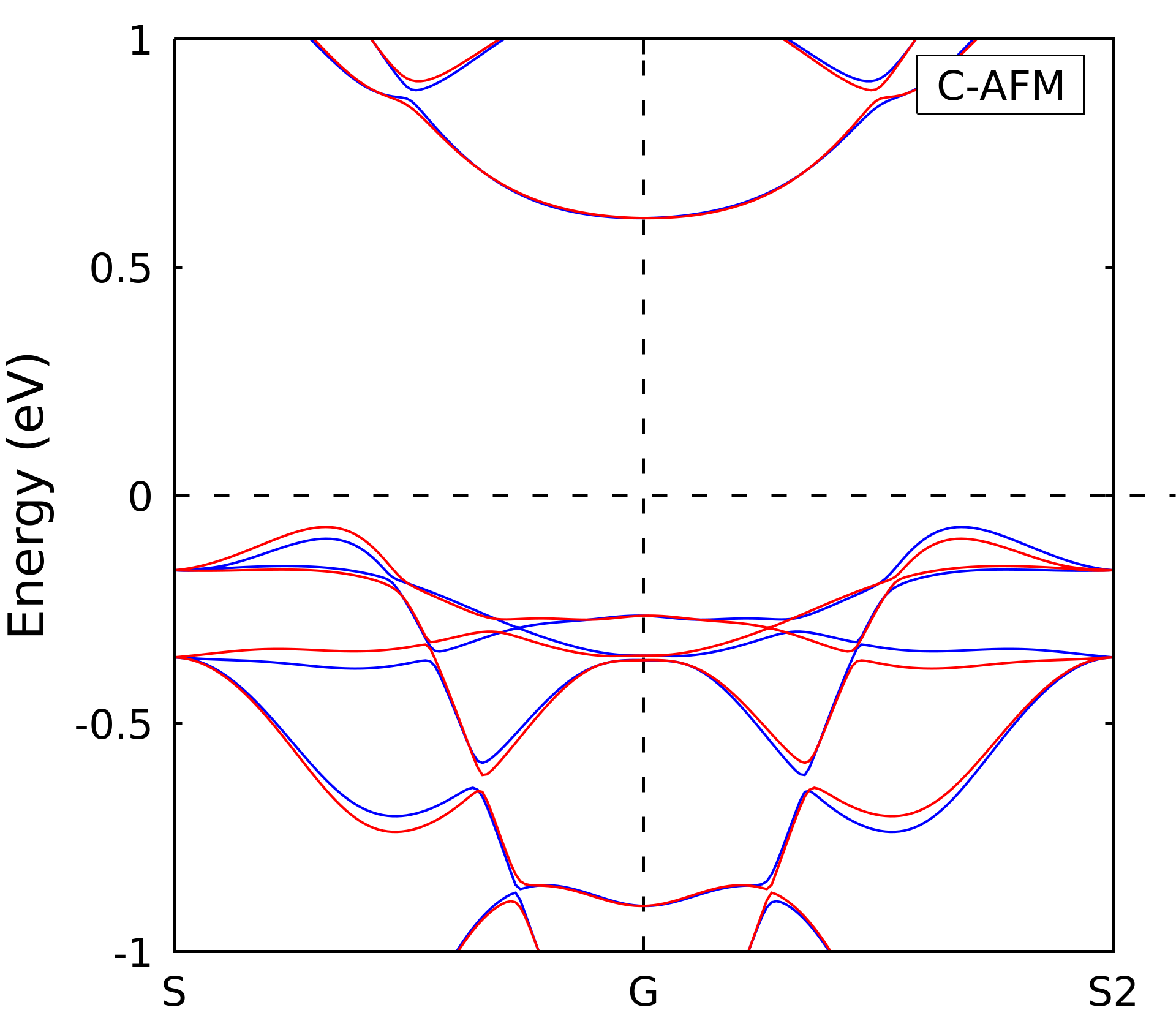}} \\ 
	  \hline
	   SrRhO$_3$ & 
	  \begin{tabular}{@{}c@{}}
	  	Ferromagnetism \\
	  	Energy (meV): \\
	  	A=123.83\\ C=106.18 \\
	  	G=124.49 \\ FM=0
	  \end{tabular} & 
	  \raisebox{-.6\height}{\includegraphics[height=4cm,width=3.5cm,keepaspectratio]{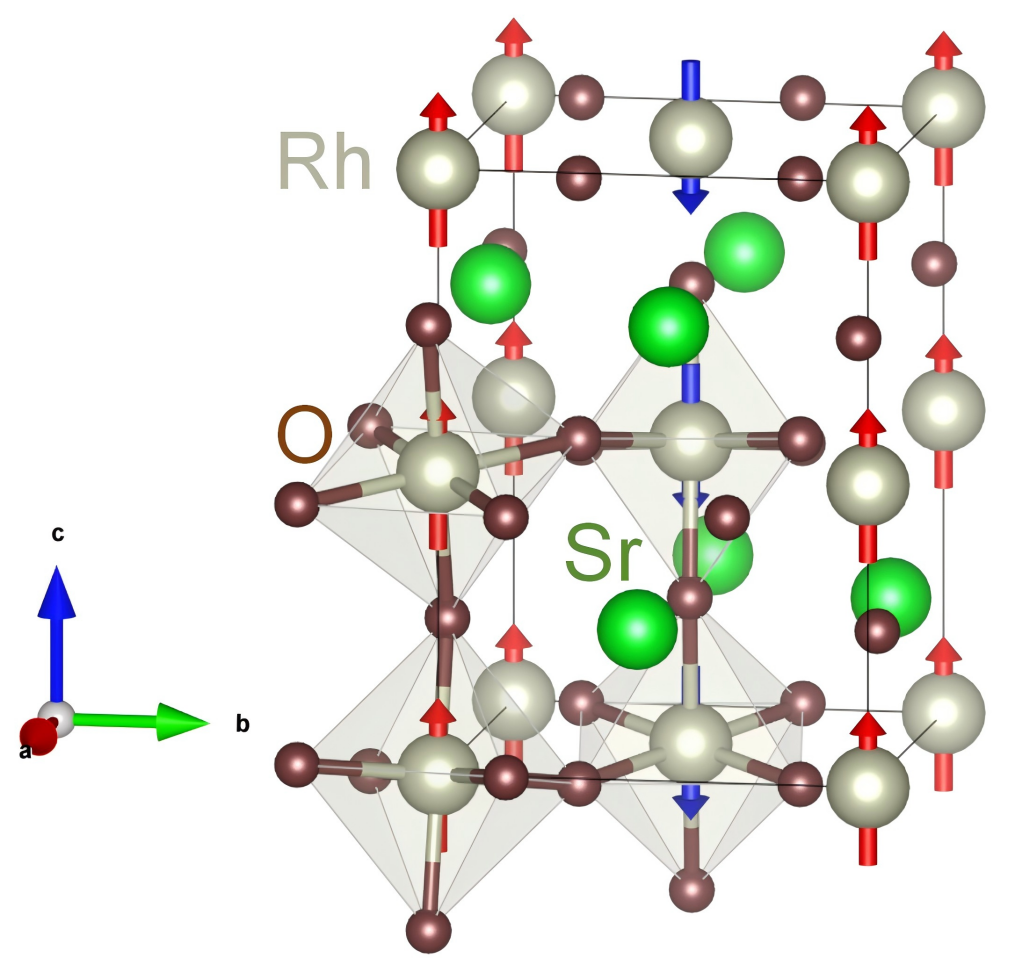}} & 
	  \raisebox{-.6\height}{\includegraphics[height=4cm,width=3.5cm,keepaspectratio]{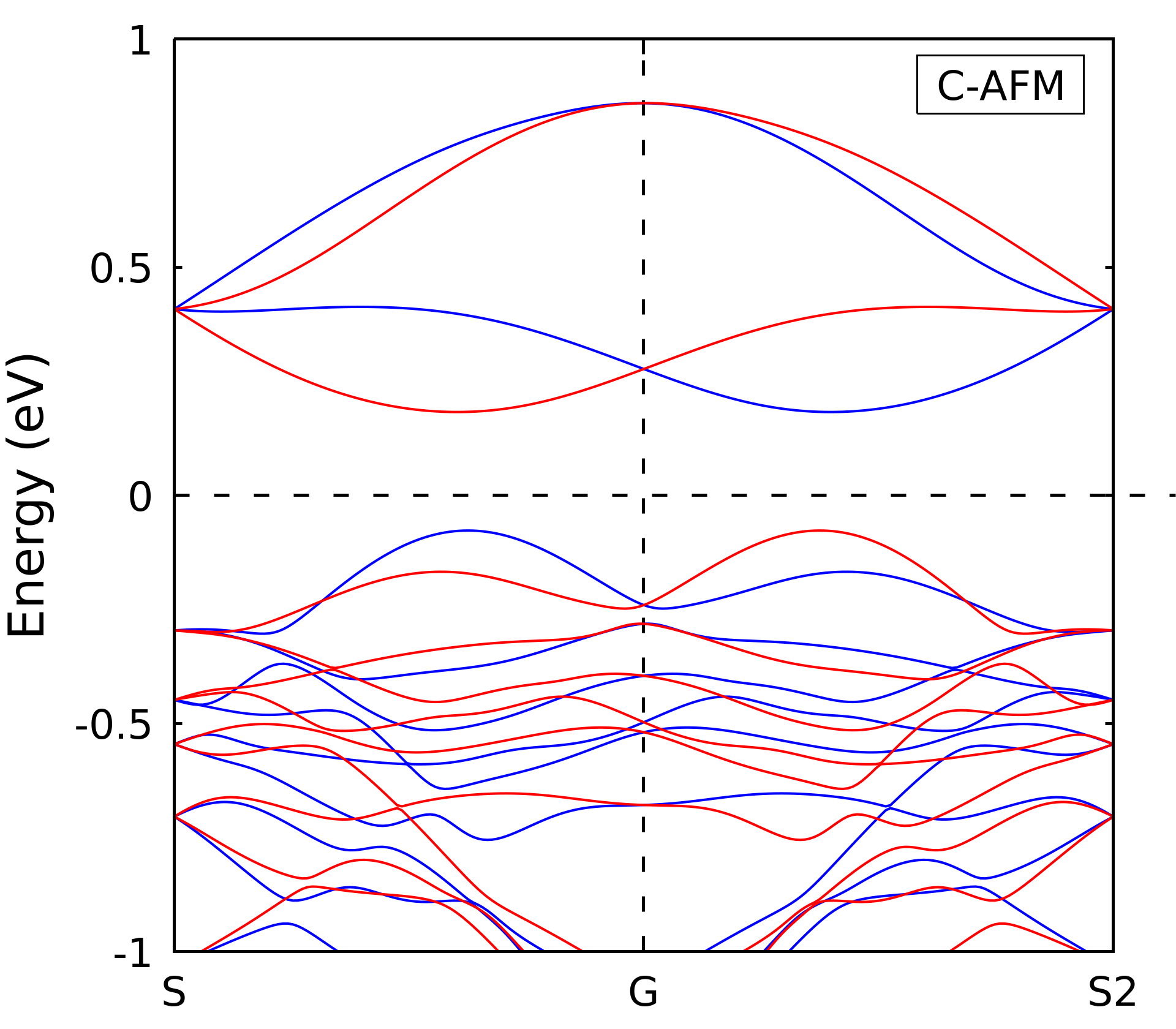}} \\ 
	  \hline
	   SrRuO$_3$ & 
	  \begin{tabular}{@{}c@{}}
	  	Ferromagnetism \\
	  	Energy (meV): \\
	  	A=257.18\\ C=477.57 \\
	  	G=1343 \\ FM=0
	  \end{tabular} & 
	  \raisebox{-.6\height}{\includegraphics[height=4cm,width=3.5cm,keepaspectratio]{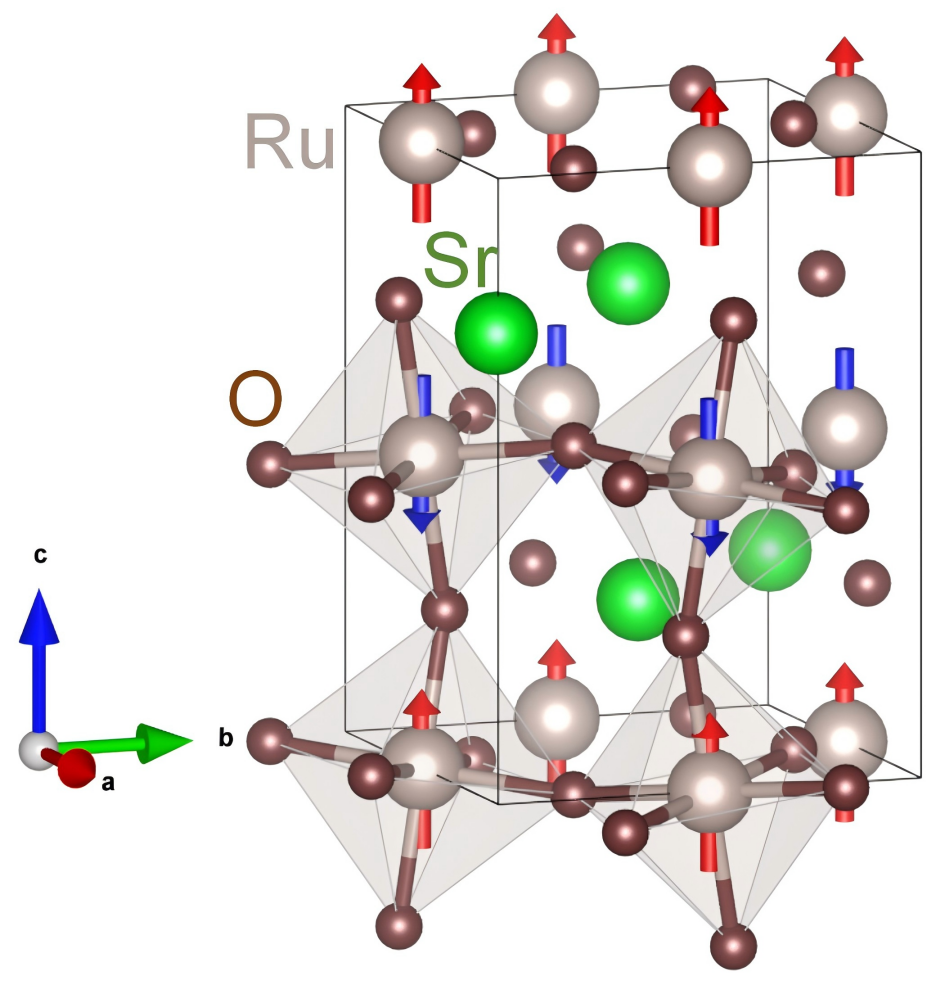}} & 
	  \raisebox{-.6\height}{\includegraphics[height=4cm,width=3.5cm,keepaspectratio]{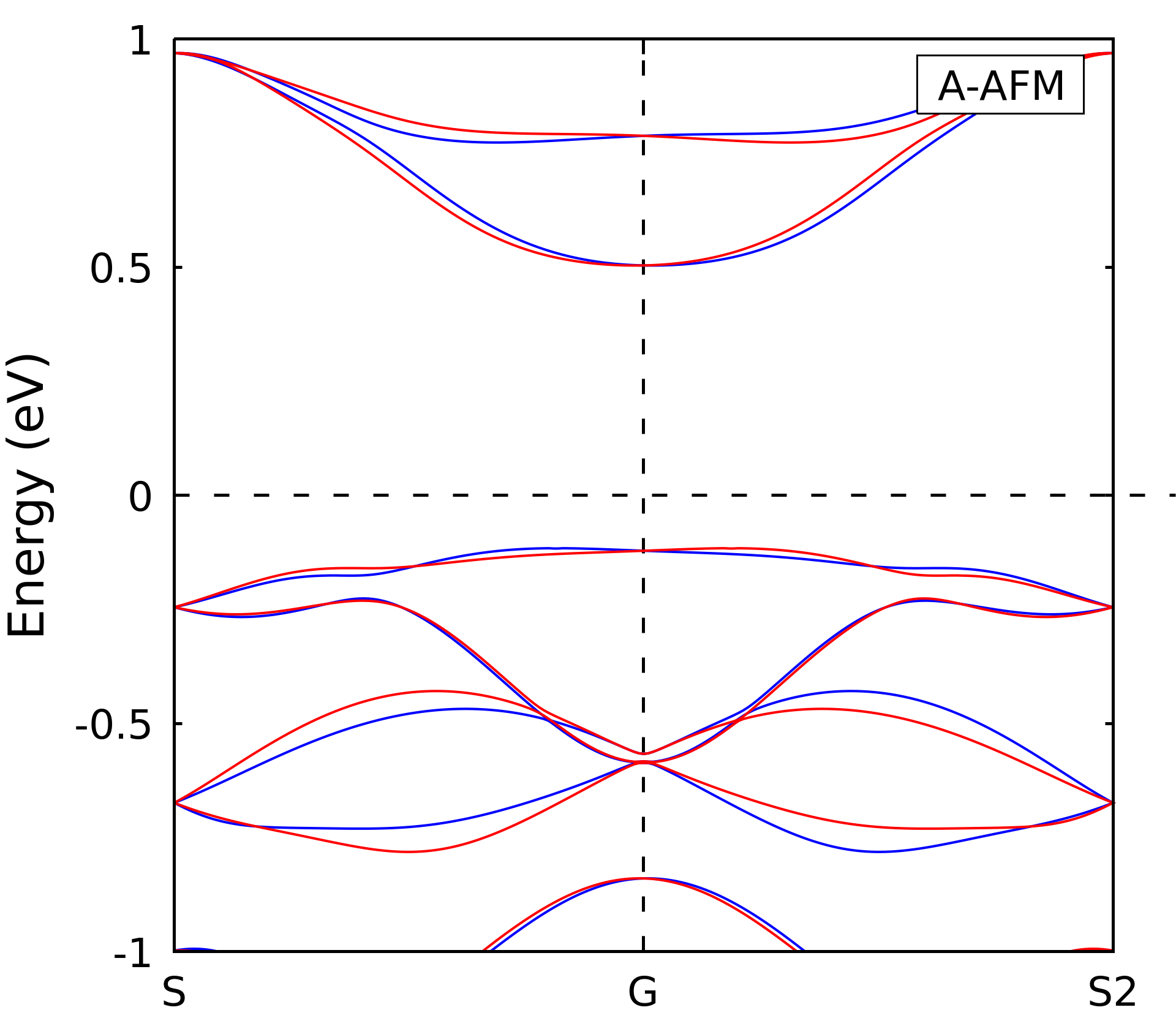}} \\ 
	  \hline
	  MnNCl$_3$ & 
	  \begin{tabular}{@{}c@{}}
	  	Antiferromagnetism \\
	  	Energy (meV): \\
	  	A=45.49\\ C=67.125 \\
	  	G=0 \\ FM=140.911
	  \end{tabular} & 
	  \raisebox{-.6\height}{\includegraphics[height=4cm,width=3.5cm,keepaspectratio]{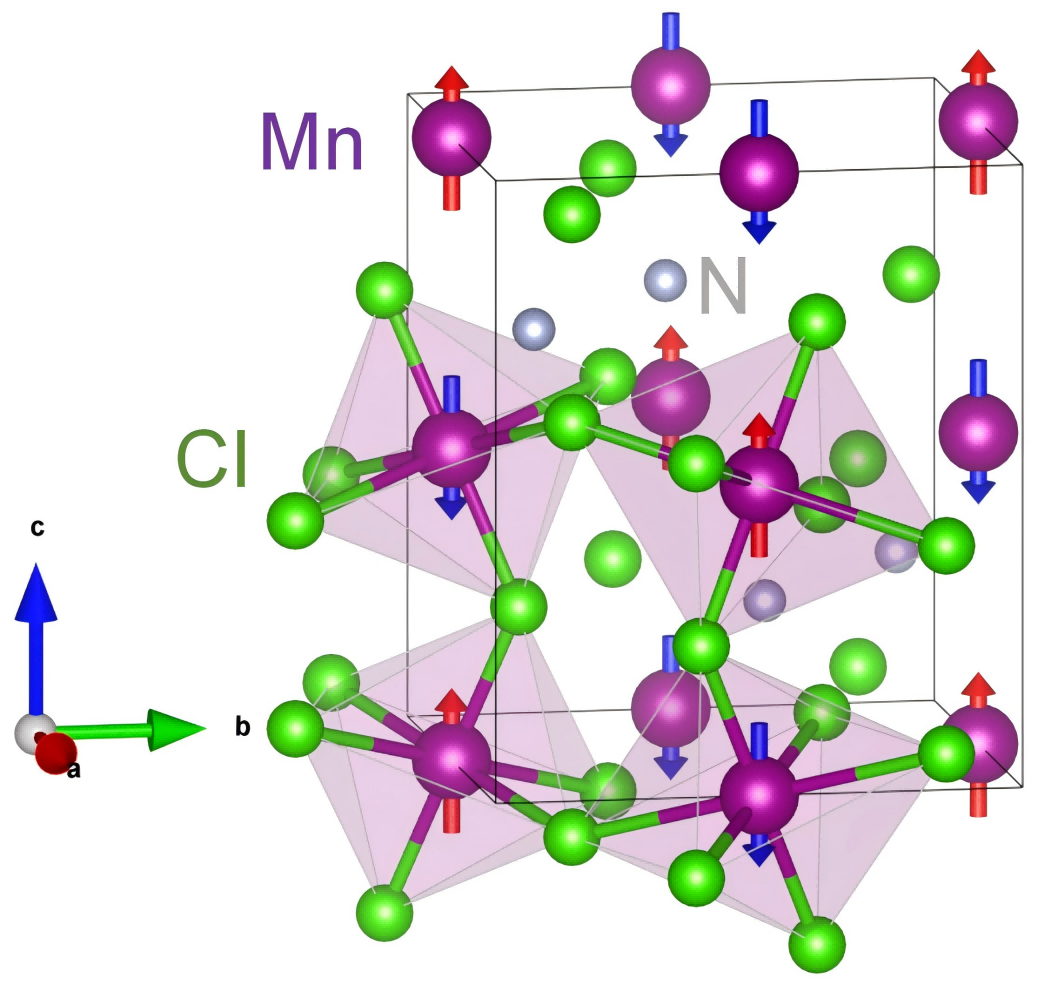}} & 
	  \raisebox{-.6\height}{\includegraphics[height=4cm,width=3.5cm,keepaspectratio]{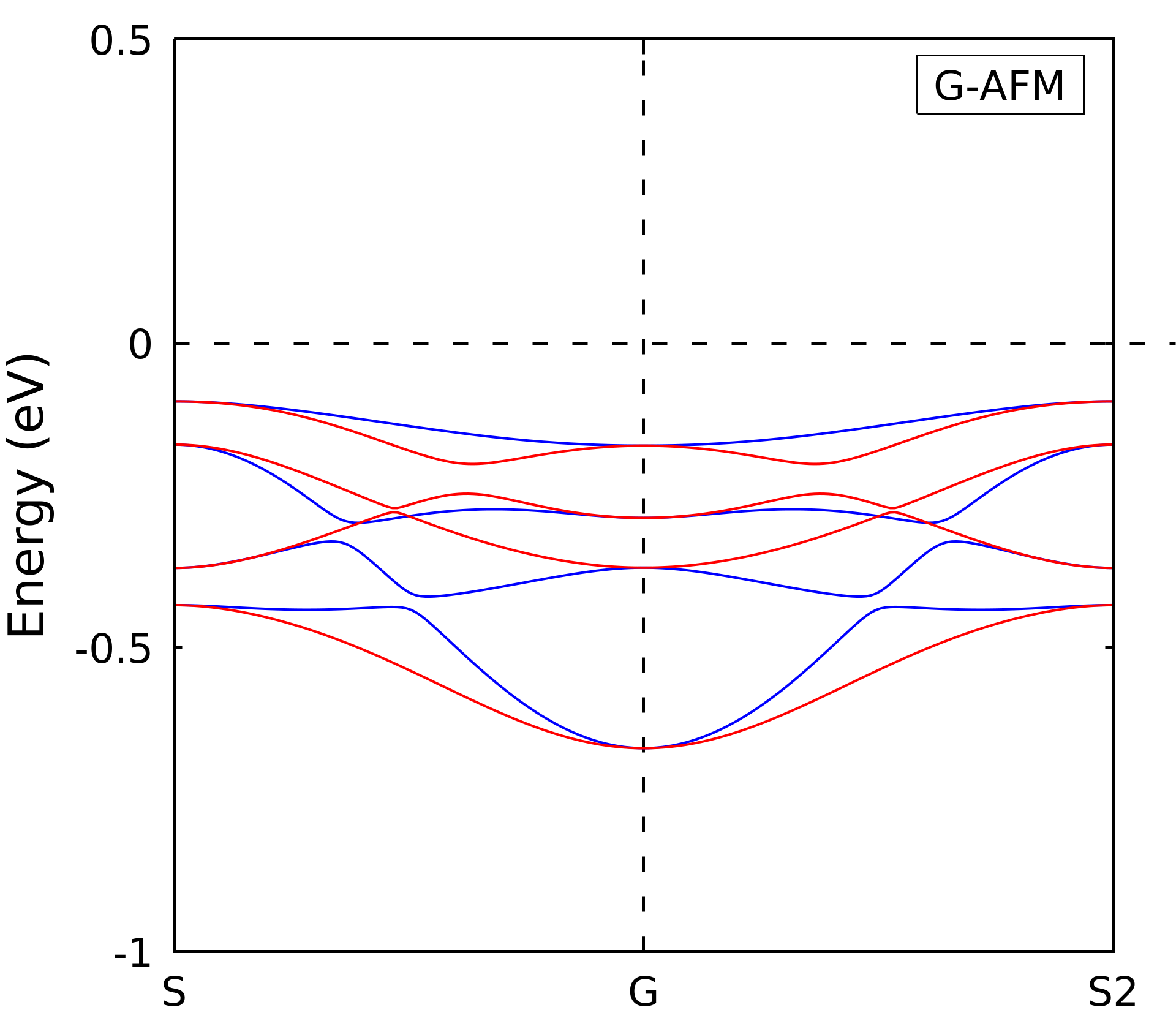}} \\ 
	  \hline
	  NaOsO$_3$ & 
	  \begin{tabular}{@{}c@{}}
	  	Antiferromagnetism \\
	  	Energy (meV): \\
	  	A=804.37\\ C=316.95\\
	  	G=0 \\ FM=1247.96
	  \end{tabular} & 
	  \raisebox{-.6\height}{\includegraphics[height=4cm,width=3.5cm,keepaspectratio]{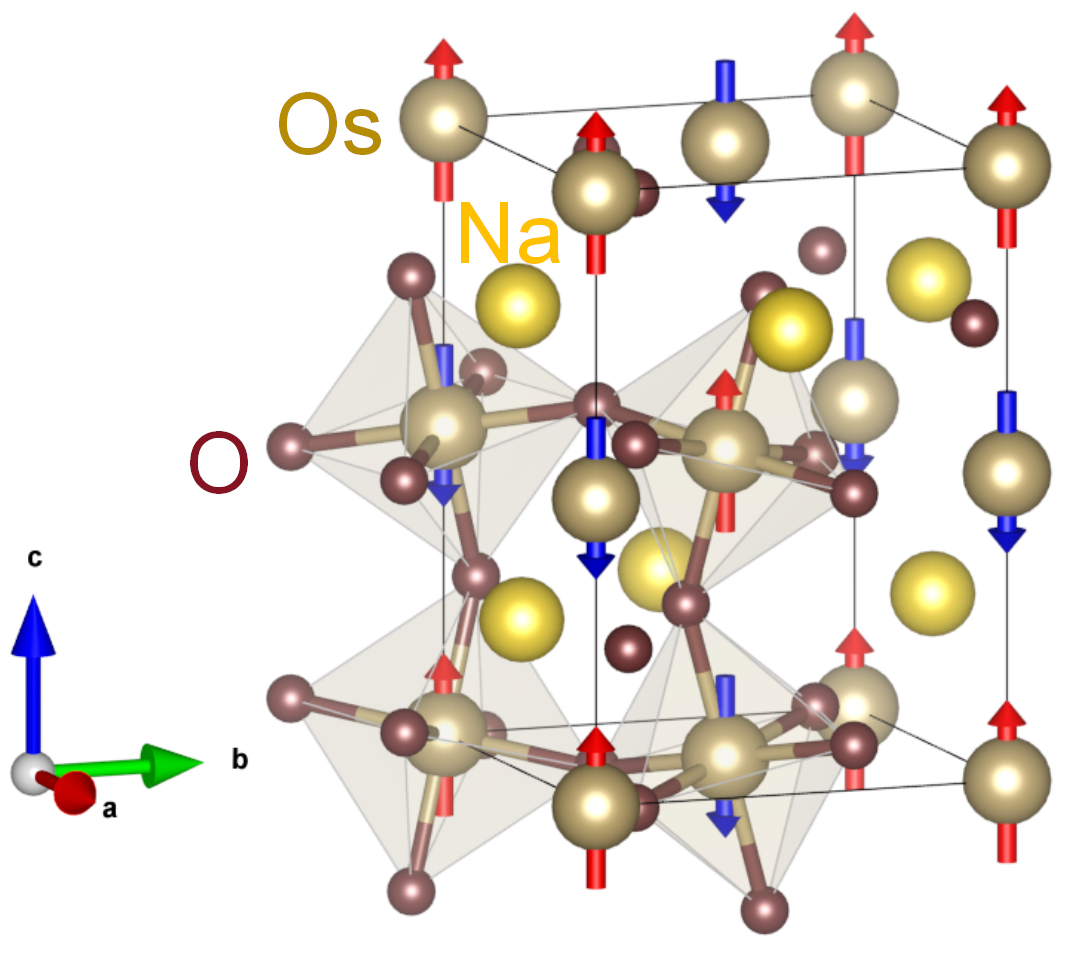}} & 
	  \raisebox{-.6\height}{\includegraphics[height=4cm,width=3.5cm,keepaspectratio]{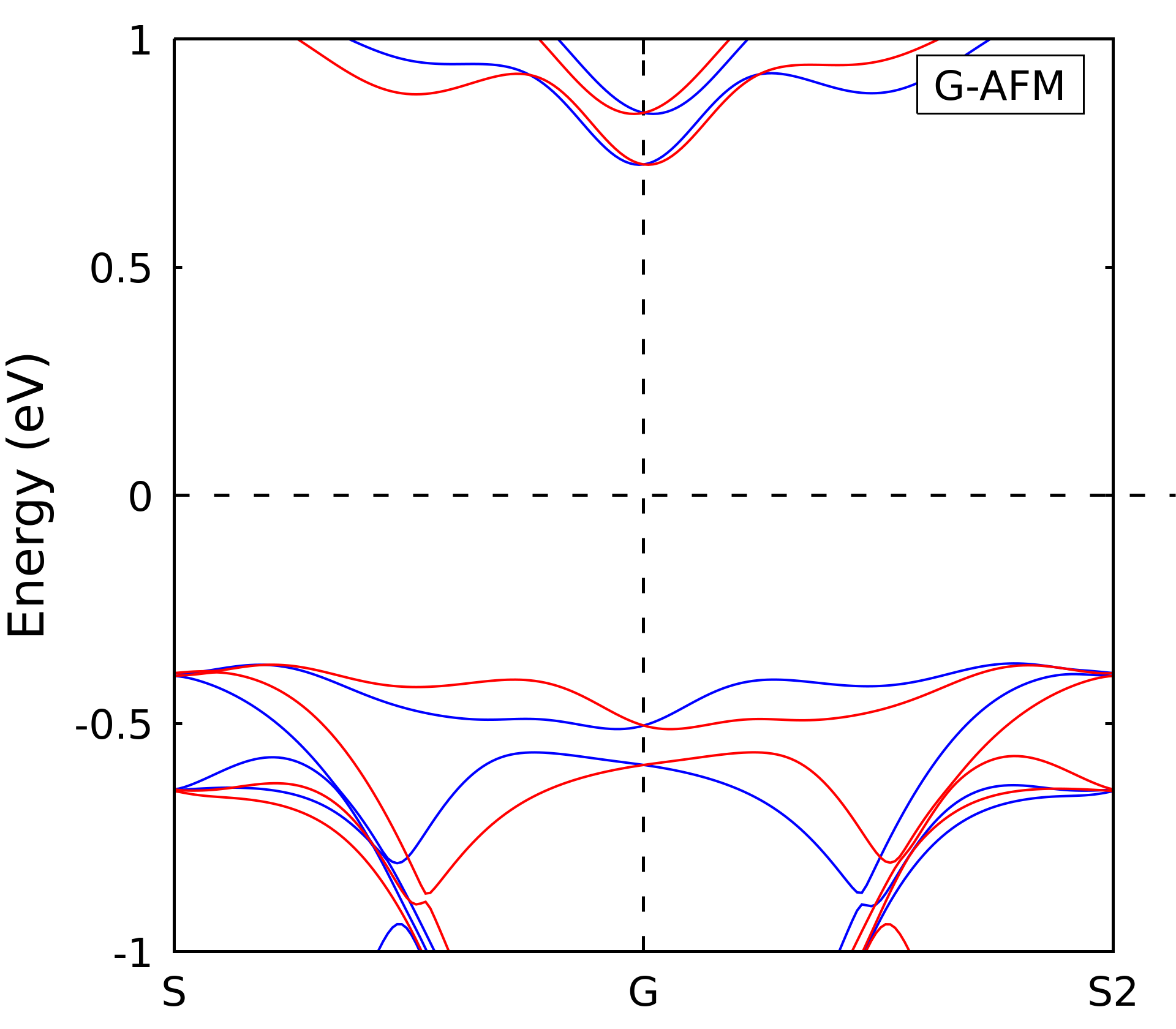}} \\ 
	  \hline
     MnTeO$_3$ & 
     \begin{tabular}{@{}c@{}}
     	Antiferromagnetism \\
     	Energy (meV): \\
     	A=44.28 \\ C=22.73 \\
     	G=0 \\ FM=74.38
     \end{tabular} & 
     \raisebox{-.6\height}{\includegraphics[height=4cm,width=3.5cm,keepaspectratio]{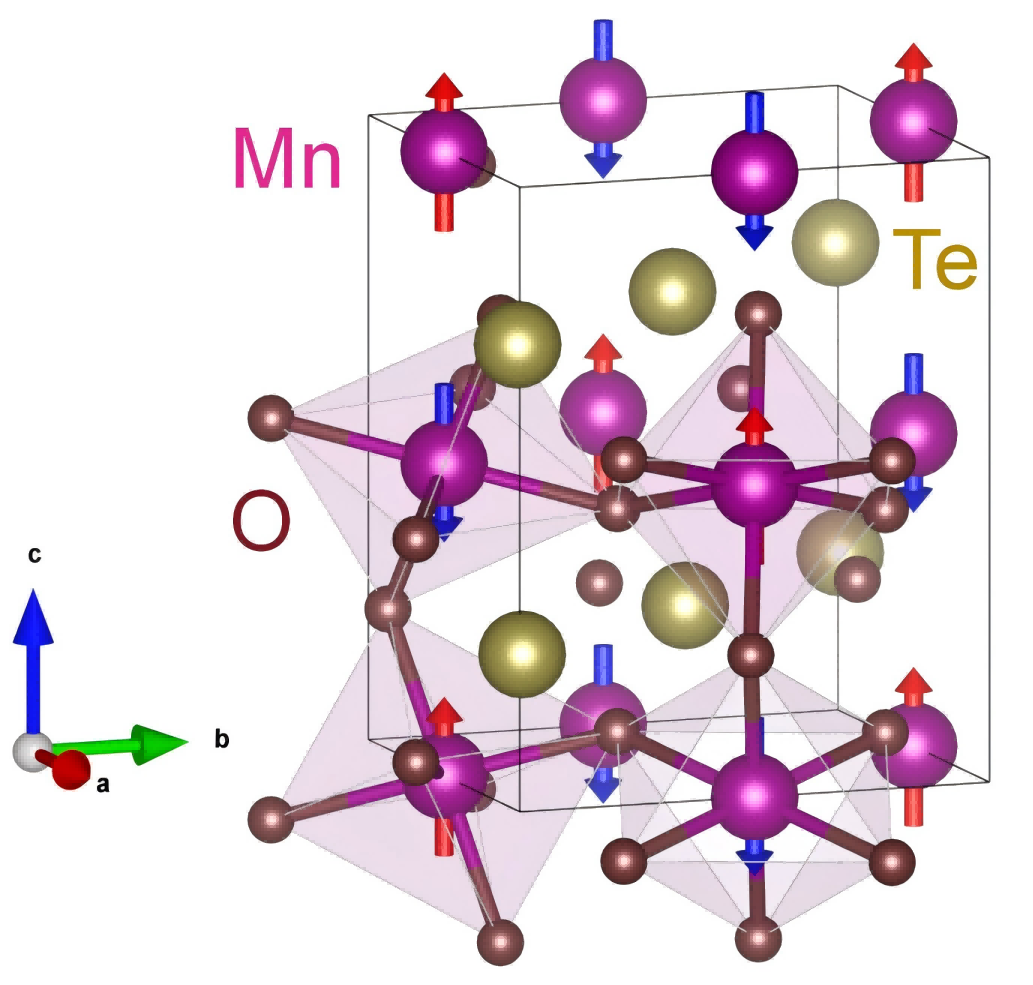}} & 
     \raisebox{-.6\height}{\includegraphics[height=4cm,width=3.5cm,keepaspectratio]{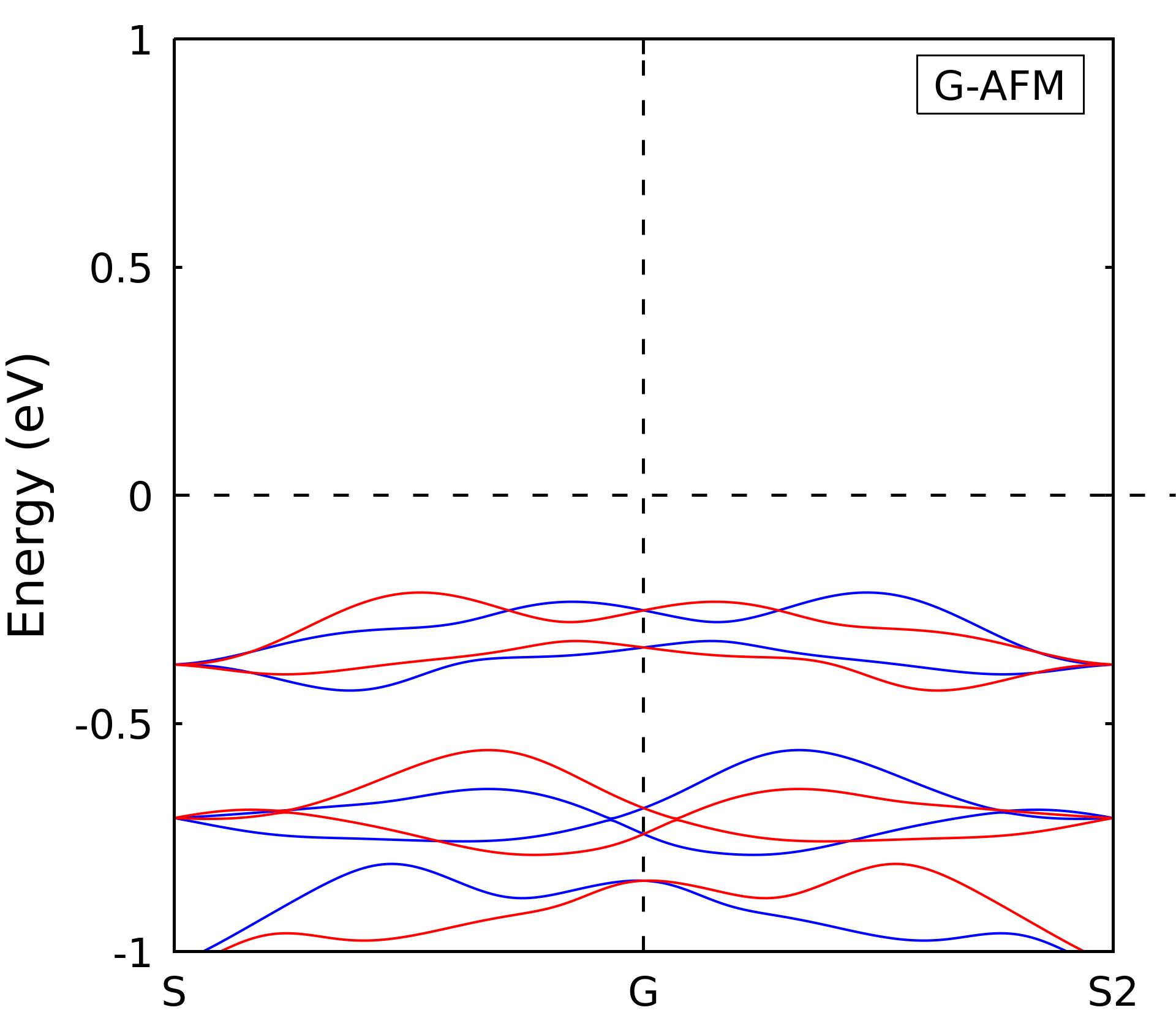}} \\ 
     \hline
     MnSeO$_3$ & 
     \begin{tabular}{@{}c@{}}
     	Antiferromagnetism \\
     	Energy (meV): \\
     	A=31.46\\ C=19.54 \\
     	G=0 \\ FM=55.23
     \end{tabular} & 
     \raisebox{-.6\height}{\includegraphics[height=4cm,width=3.5cm,keepaspectratio]{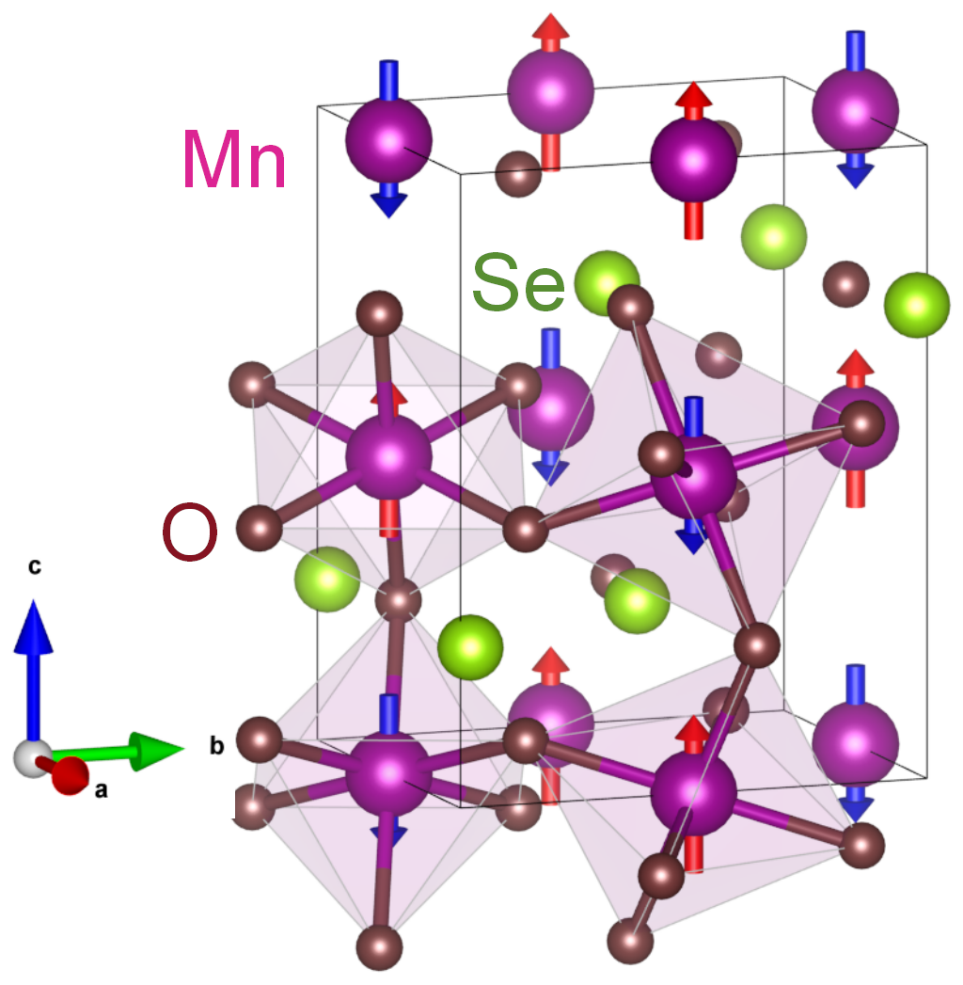}} & 
     \raisebox{-.6\height}{\includegraphics[height=4cm,width=3.5cm,keepaspectratio]{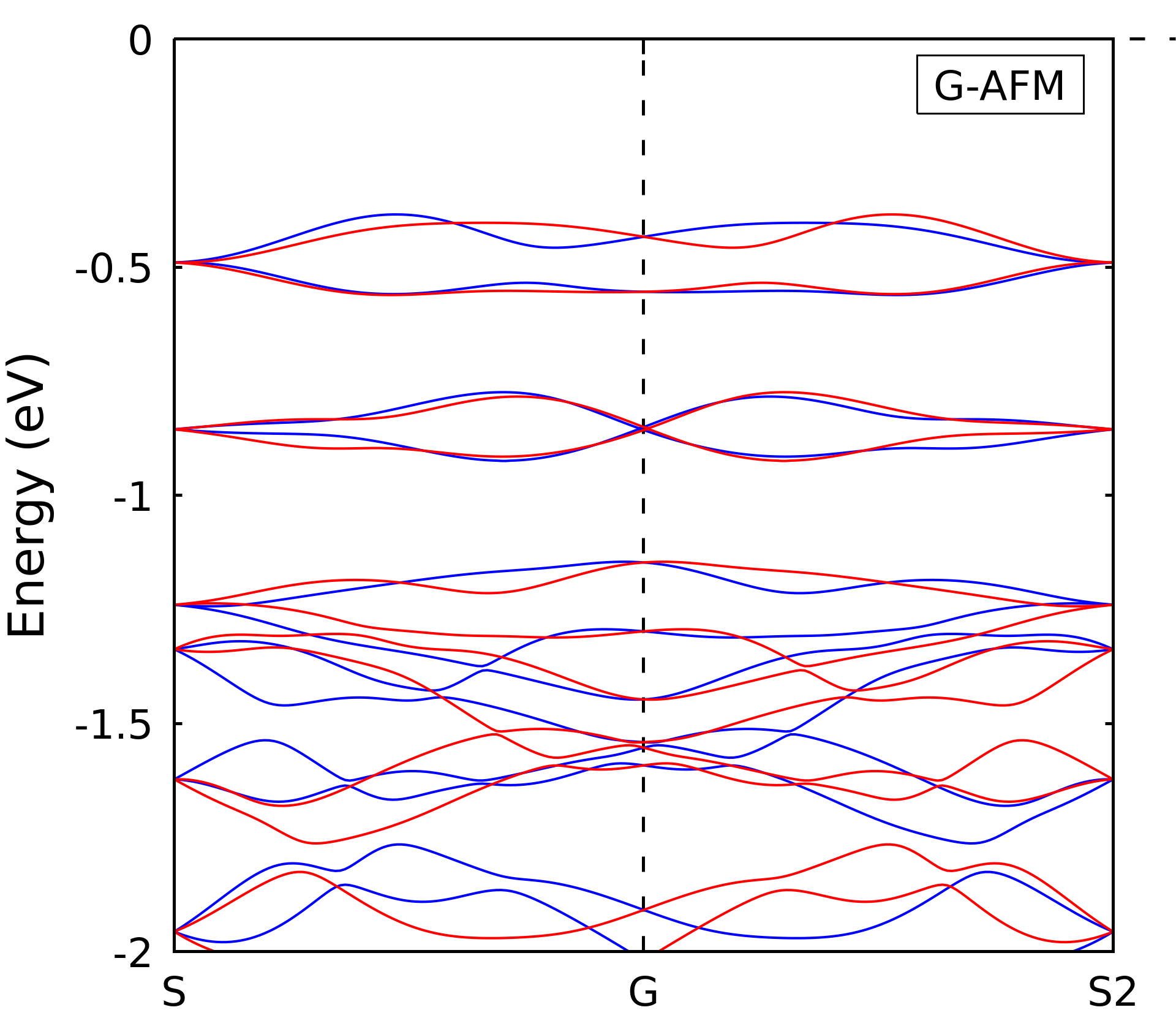}} \\ 
     \hline
     CrBiO$_3$ & 
     \begin{tabular}{@{}c@{}}
     Antiferromagnetism \\
     	Energy (meV): \\
     	A=84.44\\ C=27.04 \\
     	G=0 \\ FM=115.68
     \end{tabular} & 
     \raisebox{-.6\height}{\includegraphics[height=4cm,width=3.5cm,keepaspectratio]{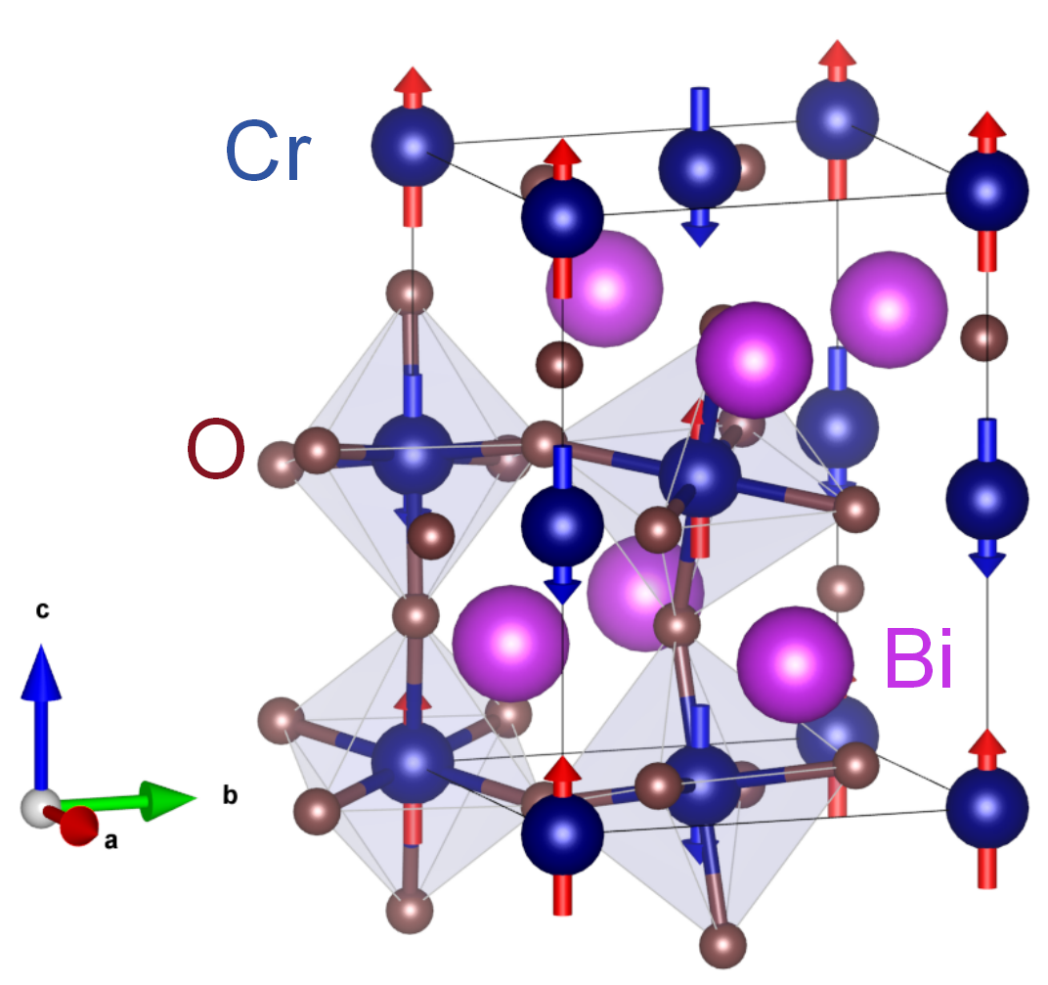}} & 
     \raisebox{-.6\height}{\includegraphics[height=4cm,width=3.5cm,keepaspectratio]{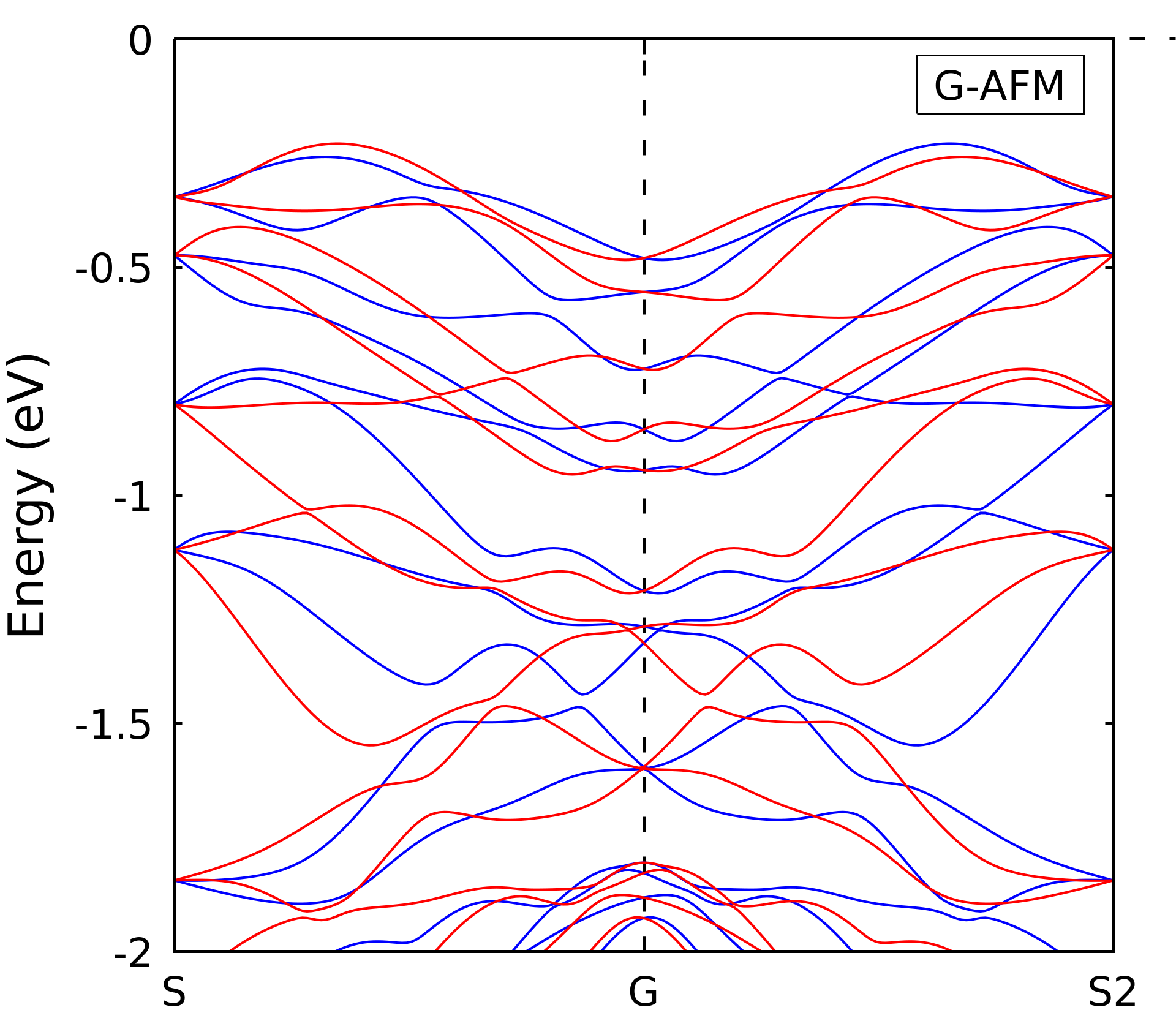}} \\ 
     \hline
	\end{longtable}

\end{document}